\documentclass[twocolumn,superscriptaddress,letter]{revtex4-2}%
\usepackage{eurosym}
\usepackage{amsbsy}
\usepackage{latexsym,epsfig}
\usepackage{graphicx}
\usepackage{dcolumn}
\usepackage{subfigure}
\usepackage{comment}
\usepackage{multirow}
\usepackage{color}
\usepackage{bm}
\usepackage{mathrsfs}
\usepackage{amssymb}
\usepackage{amsfonts}
\usepackage{amsmath}
\usepackage{xspace}
\usepackage{soul}
\usepackage{cancel}
\usepackage{float}
\usepackage{epstopdf}
\usepackage{tabularx}
\usepackage{longtable}
\usepackage[colorlinks=true, letterpaper=true, pdfstartview=FitV, linkcolor=red, citecolor=blue, urlcolor=blue]{hyperref}
\usepackage[normalem]{ulem}
\usepackage[version=4]{mhchem}
\newcommand{\be}{\begin{eqnarray}}
	\newcommand{\ee}{\end{eqnarray}}

\newcommand{\ra}{\rangle}

\begin{document}
\title{Fate of dynamical quantum phase transitions from sudden quench\\ to slow quench limit}
\author{Xiang Zhang}
\author{Fuxiang Li}
\email[Corresponding author: ]{fuxiangli@hnu.edu.cn}
\affiliation{School of Physics and Electronics, Hunan University, Changsha 410082, China}

\date{\today}

\begin{abstract}
Previous investigations of dynamical quantum phase transition (DQPT) have predominantly focused on sudden quench protocols. In this work, we systematically explore the fate of DQPT from the sudden to the slow quench regime. We establish that DQPT remains robust under slow quenching when the protocol crosses an equilibrium quantum phase transition. Conversely, we show that accidental DQPTs—which occur when the pre- and post-quench Hamiltonians reside in the same equilibrium phase—can be filtered out in the slow quench limit. This demonstrates that slow quenches can reveal the intrinsic connection between DQPT and equilibrium quantum phase transitions, thereby enabling DQPT to serve as a dynamical probe for identifying equilibrium quantum critical points. We illustrate these findings with examples from the XY chain, the Aubry-André model, and the trimer Su-Schrieffer-Heeger model. 

\end{abstract}
\maketitle

\section{Introduction}
Driven by the experimental advances in various quantum-simulator platforms such as ultracold atoms \cite{Langen} and trapped ions \cite{Leibfried}, nonequilibrium quantum dynamics \cite{Polkovnikov2011,Liu2024,Zache2019} has thrived as a central research field in physics. Intriguing out-of-equilibrium phenomena, such as many-body localization \cite{Pal,Nandkishore,Abanin,Schreiber,Smith,Choi}, quantum Kibble-Zurek mechanism \cite{Zurek2005, Polkovnikov2005, Sen2008, Barankov2008, Dziarmaga2010, Nowak2021, Kou2023, Liang2024}, time-crystal \cite{Wilczek,Sacha,Zaletel} have been theoretically predicted and experimentally verified. 
Among these, the idea of dynamical quantum phase transitions (DQPT) \cite{Heyl2013,Heyl2018dyn} has been proposed as a powerful framework to characterize quantum dynamics of  nonequilibrium many-body systems. DQPT has advanced our understanding of nonequilibrium physics in quantum many-body systems, which may lead to potential applications in quantum computing.

 Analogous to the nonanalytic behavior observed in equilibrium quantum and classical phase transitions, DQPT has been identified as the nonequilibrium counterpart in the real-time evolution of quantum systems \cite{Heyl2013,Heyl2018dyn}. A paradigmatic way to drive a quantum system far from equilibrium is through a quantum quench \cite{Polkovnikov2011}, where the system is initially prepared in the ground state of one Hamiltonian and subsequently evolved under a different Hamiltonian. In this setting, the evolution time plays the role of control parameter and can be understood as imaginary inverse temperature. A central quantity in the study of DQPT is the Loschmidt echo, defined as $\mathcal{L}(t)=|\langle\psi_0|\psi(t)\rangle|^2$ which measures the overlap between the time evolved state $|\psi(t)\rangle$ and the initial state $|\psi_0\rangle$, and its associated rate function $r(t)=-\lim_{N\to\infty}\ln\mathcal{L}(t)$, where $N$ is the number of degrees of freedom. In non-interacting models, DQPTs are signaled by nonanalyticities in the rate function $r(t)$ at critical times $t_c$, which occur when the time-evolved state becomes orthogonal to the initial state. A substantial body of work has been devoted to DQPT, both theoretically \cite{Heyl2013,Heyl2018dyn,Lang2018dynamical,Vosk2014dynamical,Bhattacharya2017inter,Huang2019dynamical,Link2020dynamical,Huang2016dynamical,Budich2016dynamical} and experimentally \cite{Jurcevic2017direct,Flaschner2018observation,Nie2020experimental,Guo20190bservation,xu2020measuring,Wang2019simulating}. The concept has also been extended to Floquet systems \cite{Yang2019floquet,Zamani2020floquet,Jafari2021floquet,Zamani2022out,Jafari2022floquet}, mixed states \cite{Bhattacharya2017mixed,Heyl2017dynamical}, finite temperatures \cite{Mera2018dynamical,Lang2018concurrence}, open quantum systems \cite{Lang2018dyn,Sedlmayr2018fate,Kyaw2020dynamical,Naji2022dissipative,Kawabata2023dynamical}, high-order topological systems \cite{Xiao2024dynamical,Maslowski2023dynamical}, and the quasiperiodic lattices \cite{Yang2017dynamical,Ye2025disentangling,Ye2024energy}. Moreover, DQPTs have been found to exhibit intriguing connections with various other phenomena, including entanglement dynamics \cite{Nicola2021entanglement,Poyhonen2021entanglement,Jafari2021floquet,Chen2025dynamical}, dynamical transition of the order parameter \cite{Heyl2014dynamical,Heyl2018dynamical,Homrighausen2017}, string-order operator \cite{Bandyopadhyay2021ob}, the work distribution function \cite{Abeling2016quantum}, and the quantum Mpemba effect \cite{Gilles2025reduced}.

Most of the investigations of DQPT have focused on the sudden quench protocols. In this context, the emergence of DQPT is highly sensitive to the specific details of the model under study.  In the relatively simple model initially explored by Heyl {\it et al.}\cite{Heyl2013}, a direct connection seemed to exist between DQPT and equilibrium quantum phase transitions—namely, DQPTs appeared to occur when the quench crossed an equilibrium phase boundary. However, subsequent studies soon revealed that such a general one-to-one correspondence does not hold \cite{Vajna2014dis,Vajna2015topo,Ye2025disentangling,Andraschko2014dyn,Karrasch2017dyn}. To date, no clear universal relationship has been established between DQPT and equilibrium criticality, and their connection remains relatively underexplored.
 
 In this work, we extend the exploration of DQPT to the slow quench protocols \cite{zhang2025uni, Puskarov2016,Sharma2016}. We establish that the occurrence of DQPT remains robust under the slow quench regime when the quench protocol crosses the equilibrium quantum phase transitions. Conversely, we show that accidental DQPTs—which occur when the pre- and post-quench Hamiltonians reside in the same equilibrium phase—can be filtered out in the slow quench limit. Therefore, the slow quenches can reveal the intrinsic connection between DQPT and equilibrium quantum phase transitions.

We illustrate our finding by taking the XY chain model, the Aubry-Andr{\'e} model and the trimer Su-Schrieffer-Heeger model as examples. For the XY chain, previous study \cite{Vajna2014dis} has reported that DQPTs can occur under sudden quench when the pre-quench and post-quench Hamiltonian belong to the same ferromagnetic phase. Our results reveal that this kind of {\color{blue}DQPT} is accidental and can be filtered out by the slow quench limit. For the Aubry-André model, Ref.~\cite{Ye2025disentangling} reported the emergence of energy-dependent {\color{blue} DQPT} during sudden quenches that begin and end within the same extended phase. We argue that these energy-dependent DQPTs are accidental in nature, as their occurrence explicitly relies on the choice of the initial state. In contrast, as the dynamical counterpart of equilibrium phase transitions, genuine {\color{blue}DQPT} should remain robust and is independent of both the initial state preparation and microscopic details of the Hamiltonian. Importantly, such state-sensitive DQPTs are also eliminated in the slow-quench limit.
 As for the trimer Su-Schrieffer-Heeger model, it hosts two distinct topological phase transitions: one accompanied by bulk gap closing and the other occurring without bulk gap closing. Moreover, we emphasize that the correspondence between slow-quench DQPT and equilibrium phase transitions is fundamentally tied to the presence of bulk gap closure. In its absence, such dynamical signatures fail to reliably capture equilibrium critical behavior. 


 This paper is organized as follows. In Sec. \ref{s2} we present a general argument of the fate of DQPT in two-band model. Models and quench dynamics are developed in Sec. \ref{s3}. We investigate three distinct models to illustrate our finding: the XY chain, the Aubry-Andr{\'e} model, and the trimer Su-Schrieffer-Heeger model and perform a comprehensive numerical study of DQPT under both sudden and slow quench protocols. We finally conclude in Sec. \ref{s4}.

\section{General Argument: DQPTs from sudden to slow quench protocols}
\label{s2}

\begin{figure}[htbp]
	\includegraphics[width=0.7\linewidth]{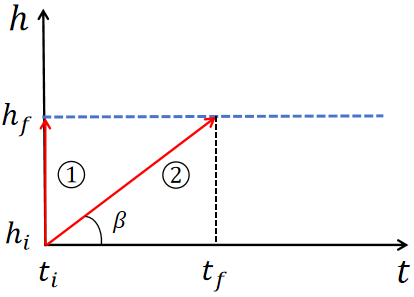}
	\caption{Schematic of the quench protocols: Sudden quench protocol \textcircled{1} and slow quench protocol \textcircled{2}. Here, for both quench protocols, the initial $h_i$ and final values $h_f$ of quench parameter $h$ are fixed. For the slow quench protocol \textcircled{2}, $h$ is changed as a function of time $t$ with quench rate $\beta$, i.e., $h_f=h_i+\beta t$.}	
	\label{p1}	
\end{figure}

We begin with a general one-dimensional two-band model. 
The Bloch Hamiltonian is of the following general form:
\begin{equation}
H_k(h)=\vec{d}_k(h)\cdot\vec{\sigma},
\label{Hk}
\end{equation}
where the Hamiltonian is parametrized by a vector $\vec{d}_k$ and the Pauli matrices $\vec{\sigma}$, and $h$ is an control parameter. For the study of nonequilibrium quantum real-time evolution and DQPTs induced by quantum quench, we prepare the system in the ground state $|\psi_k\rangle$ of the initial Hamiltonian $H_k(h_i)=\vec{d}_k(h_i)\cdot\vec{\sigma}$, then perform a sudden quench or slow quench as illustrated  in Fig.~\ref{p1}.

 Under sudden quench, a measure of the stability of time reversal is the Loschmidt echo defined by~\cite{Gorin2006dynamics,Quan2006decay,Hasegawa2021irreversibility,Jafari2017loschmidt}
\begin{equation}
\mathcal{L}(t) = \big| \langle \psi(t_i) | e^{i H(h_i) (t-t_i)} e^{-i H(h_f) (t-t_i)} | \psi(t_i) \rangle \big|^2,
\label{eq:loschmidt}
\end{equation}
where $H(h)$ is a time-independent Hamiltonian depending on a microscopic parameter $h$, whose value is assumed to change from $h_f$ during the forward time evolution to $h_i$ during the backward time evolution. $\mathcal{L}(t)$ thus is nothing but the fidelity between the state time-evolved with $H(h_i)$ and the state time-evolved with $H(h_f)$.  In this work, we are interested in the recent idea \cite{Heyl2013} to use the Loschmidt echo to study sudden quenches in closed quantum systems. We concentrate on the case where $|\psi(t_i)\rangle$ is the ground state of $H(h_i)$. In this case, the Loschmidt echo reduces to
\begin{equation}
\mathcal{L}(t) =|\mathcal{G}(t)|^2= \big| \langle \psi(t_i) | e^{-i H(h_f) (t-t_i)} | \psi(t_i) \rangle \big|^2,
\label{eq:loschmidt_reduced}
\end{equation}
where $\mathcal{G}(t)$ is the so-called Loschmidt amplitude and quantifies the deviation of the time-evolved state from the initial state \cite{Heyl2018dyn}. The Loschmidt echo \cite{Gorin2006dynamics,Quan2006decay,Hasegawa2021irreversibility,Jafari2017loschmidt} serves as a powerful tool of characterizing the quantum  decoherence  effect \cite{Zurek2001sub,Zurek2003decoherence,Cucchietti2003decoherence}, quantum chaos \cite{Gutierrez2009long,Jacquod2001golden,Cerruti2002sensitivity,Wisniacki2003short} and quantum criticality \cite{Cucchietti2004universality,tang2022dynamical}.

 We first consider the  sudden quench protocol, as shown in  Fig.~\ref{p1}(\textcircled{1}),  the parameter $h$ is switched suddenly from $h_i$ to $h_f$ at time $t=t_i$, resulting in a sudden change $\vec{d}_k(h_i)\to\vec{d}_k(h_f)$. We assume that the system
initially occupies the lower Bloch band of $H_k^i\equiv H_k(h_i)$. The associated lower band Bloch states (ground state) are denoted by $|u_k^{i,-}\rangle$ such that the initial state $|\psi(t_i)\rangle$ is a Slater determinant of all lower band Bloch states. Since lattice translation invariance is maintained at all times, the dynamics of the system can be considered separately for every lattice momentum $k$. Explicitly, we get
\begin{equation}
\begin{split}
    |\psi_k(t)\rangle&=e^{-i H_k^f (t-t_i)} |u_k^{i,-}\rangle,\\
    &=e^{iE_k^f (t-t_i)}c_{k,1} |u_k^{f,-}\rangle+e^{-iE_k^f (t-t_i)}c_{k,2} |u_k^{f,+}\rangle,
\end{split}
\end{equation}
where $E_k^f=|\vec{d}_k(h_f)|$ denotes the energy eigenvalues of $H_k^f$, i.e., $H_k^f|u_k^{f,\pm}\rangle=\pm E_k^f|u_k^{f,\pm}\rangle$ and $c_{k,1}=\langle u_k^{f,-}|u_k^{i,-}\rangle$, $c_{k,2}=\langle u_k^{f,+}|u_k^{i,-}\rangle$. Because the different momentum sectors are decoupled, ground states exhibit a factorization property. Then the Loschmidt amplitude can be expressed in a compact form:

\begin{equation}
\begin{split}
    \mathcal{G}(t)&=\prod_k\mathcal{G}_k(t)=\prod_k\langle u_k^{i,-}|\psi_k(t)\rangle\\
    &=\prod_k\left(|c_{k,1}|^2 e^{i E_k^f (t-t_i)}+|c_{k,2}|^2 e^{-i E_k^f (t-t_i)}\right).
    \label{sudden}
\end{split}
\end{equation}

{\color{blue}DQPT} occurs when the Loschmidt amplitude $\mathcal{G}(t)$ vanishes at critical times $\tau_c$ for a critical momentum $k_c$ \cite{Heyl2018dyn}. Here this can only happen if 
\begin{equation}
    |c_{k_c,1}|^2=|c_{k_c,2}|^2 \quad {\rm and} \quad \tau_{c,n}=\frac{(2n-1)\pi}{2E_{k_c}^f}-t_i,\quad n\in \mathbb{N}.
    \label{tc}
\end{equation}
{\color{blue}DQPT} hence occurs at critical momenta $k_c$ where the initial lower Bloch state is an equal weight superposition of the final Bloch states and the critical time $\tau_c$ is determined by the spectrum $E_{k_c}^f$ of the final Hamiltonian $H_{k_c}^f$. From the perspective of Eq.~(\ref{tc}), it is apparent that the survival conditions for DQPTs are not inherently linked to equilibrium quantum phase transitions. As demonstrated in Refs.~\cite{Vajna2014dis,Ye2025disentangling}, the conditions of Eq.~(\ref{tc}) can also be satisfied under sudden quench protocols when the quenching process originates and terminates within the same phases, the so-called accidental DQPTs. Therefore, the connection between DQPT and equilibrium quantum phase transitions is ambiguous when considering sudden quench protocols.


In the following, we will generalize the sudden quench protocols \textcircled{1} to slow quench protocols \textcircled{2} (Fig.~\ref{p1}) to investigate DQPTs \cite{Sharma2016,zhang2025uni,Puskarov2016}. As shown in Fig.~\ref{p1}, the quench parameter $h$ is slowly changed from $h_i$ to $h_f$ with a quench rate $\beta$, i.e., $h_f=h_i+\beta t$ during a finite time interval $[t_i,t_f]$. After the slow quenching, the subsequent temporal evolution ($t>t_f$) will be governed by the final time-independent Hamiltonian $H^f$. In the slow quench protocols, two different approaches can be adopted to quantify the Loschmidt amplitude \cite{Heyl2018dyn}. In the first approach, the Loschmidt amplitude $\mathcal{G}(t)$ and the corresponding rate function $r(t)$ are defined as
\be
\mathcal{G}(t) = \langle \psi(t_i)|\psi(t)\rangle, ~~r\left(t\right)=-\lim_{N\rightarrow\infty}\frac{1}{N}\ln |\mathcal{G}(t)|^2, 
\label{slow}
\ee
where $N$ is the number of degrees of freedom. When the quench rate $\beta$ satisfies $\beta\to \infty$, Eq.~(\ref{slow}) will return to the expression Eq.~(\ref{sudden}) of sudden quench protocols. The nonanalyticity at critical times $\tau_c$ in the rate function corresponds to DQPTs.
Here, $|\psi(t)\rangle$ {\color{blue} is} the time evolved state in  the slow quench duration  and $|\psi(t_i)\rangle$ {\color{blue}is} the initial state at initial time $t_i$. This approach has been adopted in our recent work \cite{zhang2025uni} with sufficiently slow quenching rate so that the first DQPT occurs during the slow quench duration.

In this work, all discussions will be based on the first approach. In the two-band model, the state vector $|\psi(t)\rangle$ during the slow quench process can always be formally written as a linear superposition of the instantaneous eigenstates of the time-dependent Hamiltonian $H_k(t)$:
\begin{equation}
|\psi_k(t)\rangle=p_{k,1}(t) |u_k^{-}(t)\rangle+ p_{k,2}(t) |u_k^{+}(t)\rangle.
\end{equation}
Here, $p_{k,1}(t)$ and $p_{k,2}(t)$ are the occupation amplitudes of state $|\psi_k(t) \ra$ on the two instantaneous eigenstates $|u_k^{\pm}(t)\rangle$ of time-dependent Hamiltonian $H_k(t)$, and satisfy $|p_{k,1}(t)|^2 + |p_{k,2}(t)|^2=1$.  If we start from the initial ground state $|\psi(t_i)\rangle= \prod_k|u_k^{i,-}\rangle$, the Loschmidt amplitude 
$\mathcal{G}(t)$ in Eq.~(\ref{slow}) reads
\begin{equation}
    \mathcal{G}(t)=\prod_k\left(p_{k,1}(t)\langle u_k^{i,-}|u_k^-(t)\rangle+p_{k,2}(t)\langle u_k^{i,-}|u_k^+(t)\rangle\right).
    \label{le}
\end{equation}

For simplicity, the occupation amplitudes can be rewritten as $p_{k,1}(t)=\langle u_{k}^{-}(t)|\psi_k(t)\rangle=|p_{k,1}(t)|e^{-i\theta_{k,1}(t)}$ and $p_{k,2}(t)=\langle u_{k}^{+}(t)|\psi_k(t)\rangle=|p_{k,2}(t)|e^{i\theta_{k,2}(t)}$. Meanwhile, the transition amplitudes between initial state $|u_{k}^{i,-}\rangle$ and instantaneous eigenstates $|u_k^{\pm}(t)\rangle$ can also be rewritten as
$\langle u_k^{i,-}|u_k^-(t)\rangle=|c_{k,1}(t)|e^{i\phi_{k,1}(t)}$ and $\langle u_k^{i,-}|u_k^+(t)\rangle=|c_{k,2}(t)|e^{-i\phi_{k,2}(t)}$. Therefore, the Loschmidt amplitude is 
\begin{equation}
\mathcal{G}_k(t)=\prod_k \left(|p_{k,1}||c_{k,1}|e^{i\Phi_{k,1}(t)}+|p_{k,2}||c_{k,2}|e^{-i\Phi_{k,2}(t)}\right).
    \label{le1}
\end{equation}
Here, $\Phi_{k,i}(t)=\phi_{k,i}(t)-\theta_{k,i}(t)$, $i=1,2$ and we have ignored the temporal symbols for coefficients. For sudden quench protocol \textcircled{1}, Eq.~(\ref{le1}) return to Eq.~(\ref{sudden}). Note that the magnitude of the coefficients $|c_{k,i}|$ is not related to the quench rate $\beta$, which can be seen as a constant and $|p_{k,i}|$ is closely related to the quench rate $\beta$. The existence of critical momentum $k_c$ of DQPTs requires that the magnitudes of the coefficients satisfy
\begin{equation}
    |p_{k_c,1}||c_{k_c,1}|=|p_{k_c,2}||c_{k_c,2}|.
    \label{cond}
\end{equation}
The above condition is different from the condition (\ref{tc}) (sudden quench) and the concrete values ($|p_{k_c,i}|$ and $|c_{k_c,i}|$) of the magnitudes for coefficients depend on models.

Before we discuss the specific models in next section, we want to first  understand the fate of DQPTs under slow quench by making a qualitative discussion of  Eq.~(\ref{cond}). One can obtain the values of $p_{k, 1}(t)$ and $p_{k, 2}(t)$ by numerically solving the Schrodinger equation with time-dependent Hamiltonian: 
\begin{equation}
   i\frac{d|\psi_k(t)\rangle}{dt}=H_k(t)|\psi_k(t)\rangle.
   \label{se}
\end{equation}
 If one linearly quenches the parameter $h$ in the two-band model (\ref{Hk}), the time-dependent Hamiltonian can be written in the following form 
\begin{equation}
    H_k(t)=\begin{pmatrix}
        h_k(t) & \Delta_k \\
        \Delta^*_k & -h_k(t)
   \end{pmatrix},
   \label{lz}
\end{equation}
after, if necessary, making a  transformation of representation. A specific example can be found in Sec.~\ref{sec:XY} for the XY chain model. Here, $h_k(t)=h_i+\beta t$, and $\Delta_k$ is the coupling between the two energy levels. 
This is the famous Landau-Zener-St${\rm \ddot{u}}$ckelberg-Majorana problem \cite{Landau1932theorie,Zener1932non,Stueckelberg1932theorie,Majorana1932atomi}, or simply called Landau-Zener problem.  Landau-Zener problem provides an analytical solution for the transition probability from the ground state at initial time $t\rightarrow -\infty$ to the excited state at final time $t\rightarrow \infty$ 
\begin{equation}
    P=e^{-\frac{\pi\Delta_k^2}{\beta}}.
\end{equation}
For general cases of time evolution, the above Landau-Zener formula can serve as a good approximation. 

When the system is quenched across the quantum phase transition, there always exists a $k_0$ point such that  $\Delta_{k_0}=0$, so that two energy levels simply cross with each other with no transition between the two levels. At this $k_0$ point, if starting from the ground state, the system will finally go to the upper level, and the transition amplitude $|p_{k_0,2}|=1$. For other $k$ points away from the critical point $k_0$, $\Delta_k$ is finite, and the transition probability $|p_{k_0, 2}|$ decreases towards zero with decreasing quench rate $\beta$. In the thermodynamic limit ($N\to\infty$), momentum is continuous and thus we can always find a special point $k_c$ such that the condition (\ref{cond}) is satisfied, which constitutes a necessary condition for the DQPTs. 

One the other hand, if we do not quench system across the quantum phase transition, but within the same phase, the system is always gapped, and $\Delta_k$ is always nonzero. As long as the quench rate is sufficiently slow, i.e., very small $\beta$, the transition amplitude $p_{k, 2}$ would be always very small for each $k$, so that the condition (\ref{cond}) for DQPTs can not be satisfied, and no DQPTs can occur. 
Therefore, slow quench dynamics can rule out accidental DQPTs that appear within the same phase under sudden quench. 

There is an alternative approach for the Loschmidt amplitude $\mathcal{G}(t)$, which is defined as 
\begin{equation}
    \mathcal{G}(t)=\langle\psi(t_f)|\psi(t)\rangle,
\end{equation}
 in which $|\psi(t)\rangle$ is the time evolved state governed by the final time-independent Hamiltonian $H^f$ after the slow quench duration, i.e., $|\psi(t)\rangle=e^{-iH^f(t-t_f)}|\psi(t_f)\rangle$ and  $|\psi(t_f)\rangle$ is the  state vector at the end of the slow quench duration at time $t_f$ \cite{Puskarov2016,Sharma2016}. Note that, in the slow quench protocols, the second approach is equivalent to preparing a different initial state, which renders the  comparison with its sudden qunech counterpart unreasonable. Therefore, we will restrict our discussions within the first approach.

\section{Model and Quench dynamics}
\label{s3}
In this section, we investigate three different models: the XY chain, the Aubry-André model, and the trimer Su-Schrieffer-Heeger model, and numerically study  the DQPT under both sudden and slow quench protocols, within the same phase and across the phase transitions. We will explore the fate of DQPT from the sudden to the slow quench regime. We will demonstrate that DQPTs remain robust under slow quenching when the protocol crosses an equilibrium quantum phase transition. In contrast, accidental DQPTs, which occur when the pre- and post-quench Hamiltonians reside in the same equilibrium phase, will disappear in the slow quench limit. 


\subsection{XY Chain \label{sec:XY}} 
We begin with the spin-1/2 quantum XY chain under a uniform transverse field (homogeneous for each spin). The Hamiltonian of the transverse-field XY chain with nearest-neighbor interaction is given by
\begin{equation}
	 H=-\sum_{j=1}^N\left[\frac{1+\gamma}{2}\sigma_j^x\sigma_{j+1}^x+\frac{1-\gamma}{2}\sigma_j^y\sigma_{j+1}^y+h\sigma_j^z\right],
\end{equation}
where $N$ counts the number of spins, $\sigma_j^i$ $(i=x,y,z)$ are the Pauli matrices acting on the $j$th spin, $\gamma$ represents the strength of anisotropy, and $h$ measures the strength of the transverse field. The system reduces to the isotropic XY chain for $\gamma\to 0$ and {\color{blue}to} the Ising chain for $\gamma\to 1$.

This model is exactly solvable by the Jordan-Wigner transformation, which maps a system of spin 1/2 to a system of spinless free fermions \cite{Jordan1928paulische,Dziarmaga2005}. After the Jordan-Wigner transformation, the XY chain Hamiltonian can be rewritten as
\begin{equation}
	\begin{split}
		H&=-\sum_{j=1}^N\left(c_j^\dagger c_{j+1}+\gamma c_j^\dagger c_{j+1}^\dagger+{\rm H.c.}\right)\\
&+h\sum_{j=1}^N\left(2c_j^\dagger c_j-1\right),
	\end{split}
\end{equation}
where H.c. denotes the Hermitian conjugate. After the Fourier
transformation of the fermionic operators, we obtain the momentum-space representation of the Hamiltonian,
\begin{equation}
	H=\sum_k\psi_k^\dagger H_k\psi_k
    \label{xyha}
\end{equation}
where $\psi_k=(c_k,c_{-k}^\dagger)^T$ are Nambu spinors and $H_k=\vec{d_k}\cdot\vec{\sigma}=\gamma \sin k\sigma_y+(h-\cos k)\sigma_z$. The energy is $E(k)=\pm\sqrt{\gamma^2 \sin^2k+(h-\cos k)^2}$. The system has  quantum critical points at $h=\pm1$. For the $h=1$, the gap closing occurs at $k=0$, while for $h=-1$, it occurs at $k=\pm\pi$. This model exhibits two phases, $|h|<1$ for ferromagnetic phase and $|h|>1$ for paramagnetic phase. 

 Previous studies of DQPTs under sudden quench in the XY model have shown that, even when the pre- and post-quench Hamiltonians are in the same ferromagnetic phases,   
 DQPT can still occur \cite{Vajna2014dis}. Here, we extend the investigation of DQPT from sudden quench regime to the slow quench regime, and show that under slow quench regime, accidental DQPTs which occur when the pre- and post-quench Hamiltonians reside in the same equilibrium phase, will disappear. Explicitly, we  implement two distinct quench protocols within the same ferromagnetic phase with fixed $\gamma$: (i) sudden quench protocol, where the field $h$ is abruptly switched from $0$ to $0.9$ at time $t=0$; (ii) slow linear quench protocol, with $h(t)=h_i+ \beta t$, where the initial and final values $h_i=0$ and $h_f=0.9$ which are the same as those of the sudden quench protocol.

Under the slow quench, the time-dependent Hamiltonian $H_k(t)$ is
 \begin{equation}
     H_k(t)=\begin{pmatrix}
        \beta t-\cos k & -i \gamma\sin k \\
        i \gamma\sin k & -\beta t+\cos k
   \end{pmatrix}.
 \end{equation}
 Compared to Eq.~(\ref{lz}), $h_k(t)=\beta t-\cos k $ and $\Delta_k=-i\gamma \sin k$. At $k_0=0$ or $\pi$, the off-diagonal element cancels $\Delta_k=0$  so that the two energy levels cross with each other. Therefore, if the system is quenched across the quantum phase transition, i.e., $t$ is taken from $0$ to a final time $t_f>1/\beta$,  the transition amplitude reaches $|p_{k_0,2}|=1$. For other $k$ points away from the critical point, $\Delta_k$ is finite and the transition probability decreases towards zero with decreasing $\beta$. Therefore, we can always  find a special point $k_c$ such that condition (\ref{cond}) is satisfied, which constitutes a necessary condition for the DQPTs.

\begin{figure}[htbp]
	\includegraphics[width=1\linewidth]{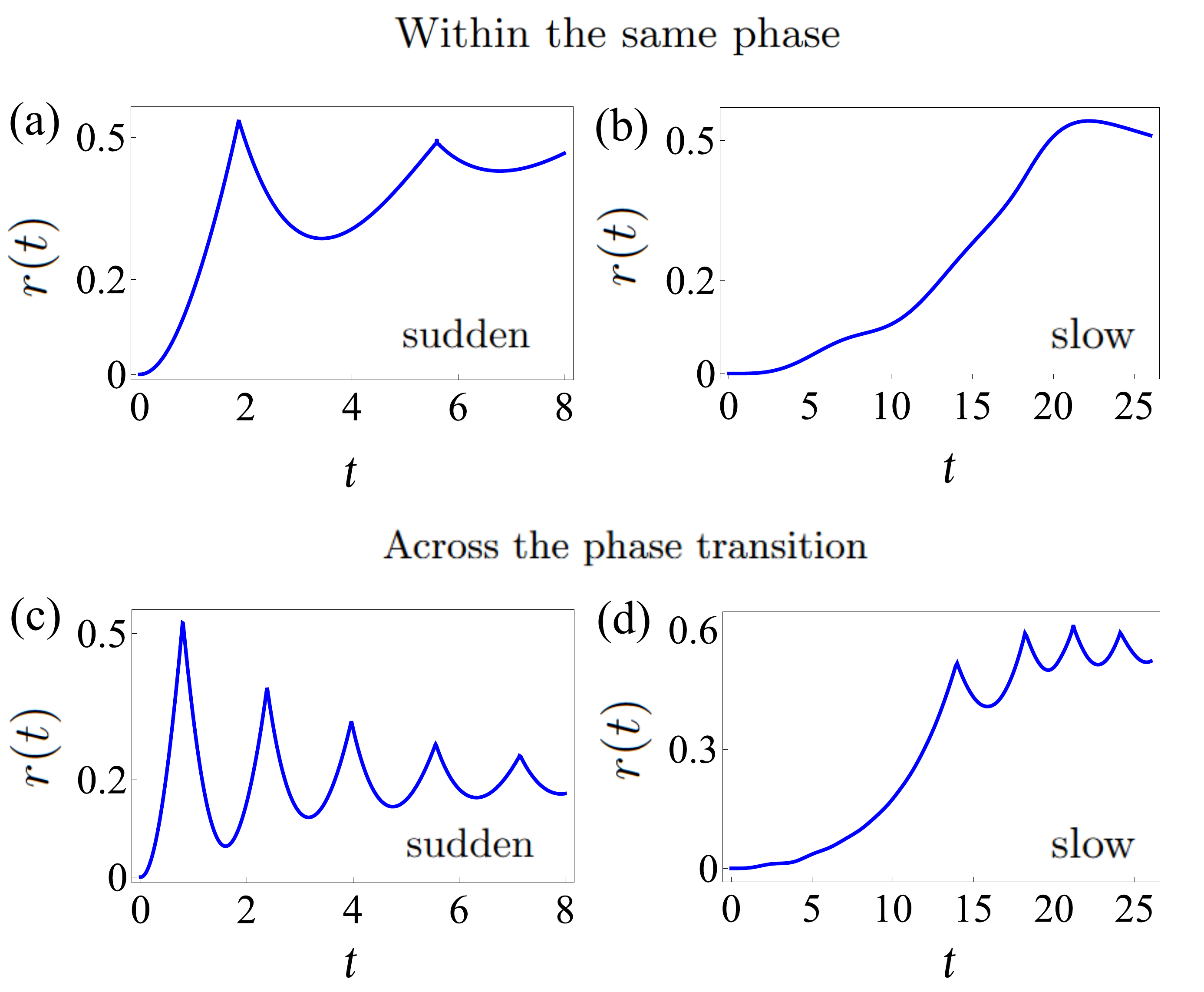}
	\caption{DQPTs in the XY chain under sudden and slow quench protocols when the system is quenched within the same ferromagnetic phase and across the phase transition from ferromagnetic phase to paramagnetic phase, respectively. (a) and (b) The rate function $r(t)$ of the Loschmidt amplitude under sudden and slow quench protocol within the same phase. (a) Parameter $h$ is suddenly changed from $h_i=0$ to $h_f=0.9$ at time $t=0$ while $\gamma=0.3$.  (b) The quench parameter $h$ is slowly changed as $h(t)=h_i+\beta t$ with the quench rate $\beta=0.05$. (c) and (d) The rate function $r(t)$ of the Loschmidt amplitude under sudden and slow quench protocol across the equilibrium phase transition. (c) $h$ is suddenly changed from $h_i=0$ to $h_f=2$ at time $t=0$. (d) The quench parameter $h$ is slowly changed as $h(t)=h_i+\beta t$ with same quench rate $\beta=0.05$}.	
	\label{xy1}	
\end{figure}

We now begin to study DQPT under slow quench by numerically solving Eq.~(\ref{se}).  As shown in Fig.~\ref{xy1}(a), the nonanalyticities in the rate function $r(t)$ appear that correspond to the occurrence of DQPTs under a sudden quench even when the pre-quench and post-quench Hamiltonians belong to the same equilibrium phase, which is consistent with the results from Ref.~\cite{Vajna2014dis}.  In stark contrast, under the slow quench protocol with the same initial and final value of control parameters, DQPTs disappear that are manifested as the absence of the nonanalyticity, as shown in Fig.~\ref{xy1} (b). Here, the quench rate is $\beta=0.05$. These results indicate that the DQPTs observed in the corresponding sudden quench are accidental in nature and can be filtered out by adopting a sufficiently slow quench protocol. 

For comparison, we also investigate DQPT under quench protocols across the equilibrium quantum phase transitions. Specifically, we also consider two different quench protocols with fixed $\gamma$: (i) sudden quench protocol, where the field $h$ is abruptly switched from $0$ to $2$ at time $t=0$; (ii) slow linear quench protocol, with $h(t)=h_i+ \beta t$, where the initial and final values $h_i=0$ and $h_f=2$ which are the same as those of the sudden quench protocol. Here, we choose the same quench rate $\beta=0.05$. As shown in Fig.~\ref{xy1} (c) and (d), we reveal that DQPT is robust even under slow quench protocols, as shown in Fig.~\ref{xy1} (d). Therefore, we can establish a intrinsic connection between DQPT and equilibrium quantum phase transitions under the slow quench protocol, thereby enabling DQPT to serve as a dynamical probe for identifying equilibrium quantum phase transition.

\begin{figure*}[htbp]
	\includegraphics[width=0.9\linewidth]{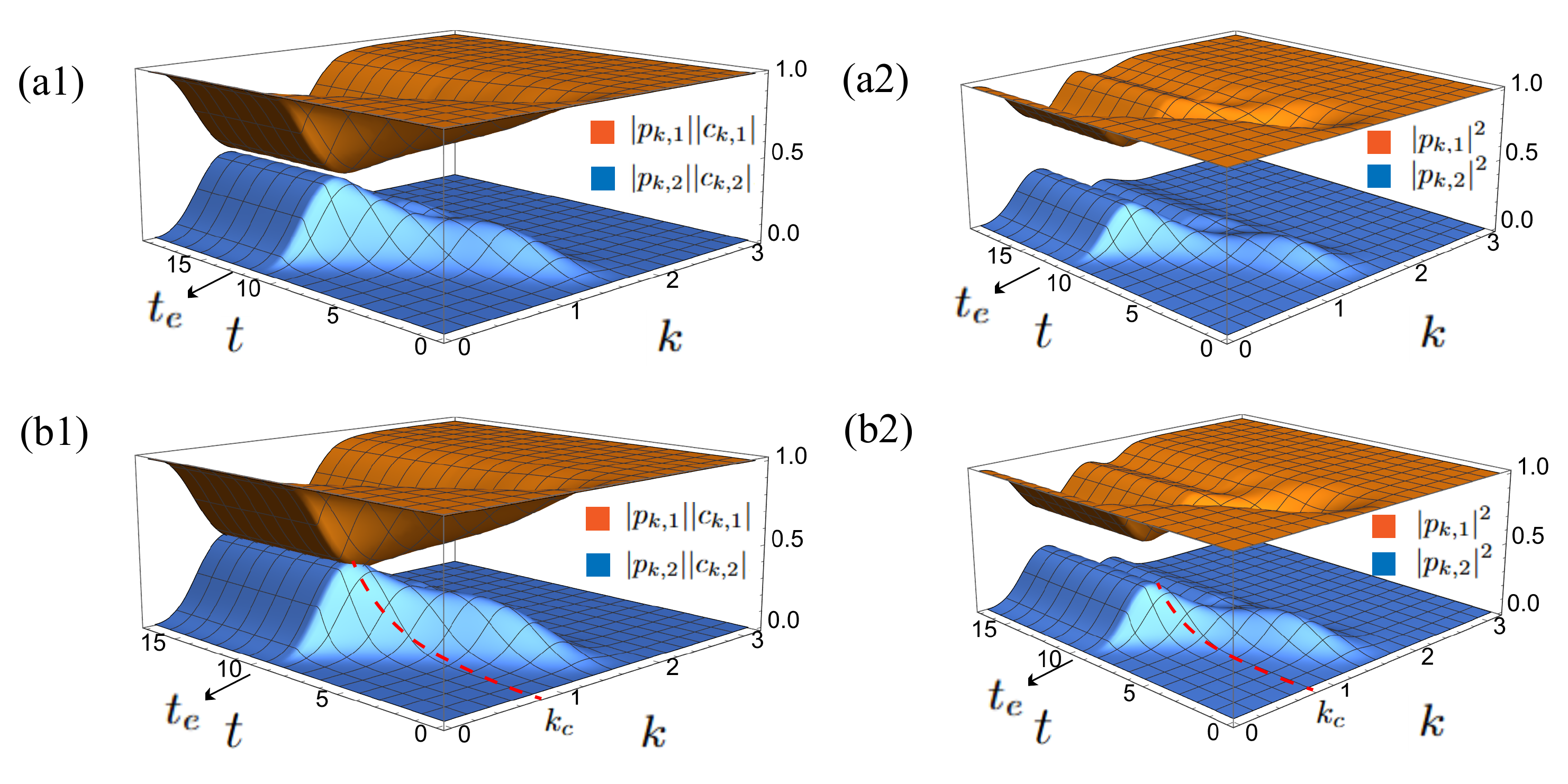}
	\centering
	\caption{The condition (\ref{cond}) for quench rate $\beta=0.08$ (a1) and $\beta=0.1$ (b1) in the XY chain within the same phase. Occupation amplitudes $p_{k,1}$ and $p_{k,2}$ for quench rate $\beta=0.08$ (a2) and $\beta=0.1$ (b2). $t_e$ represents the ending time of slow quench protocols. After $t_e$, the subsequent temporal evolution will be governed by the final time-independent Hamiltonian. In (b1) and (b2), the red dotted lines represent the time evolution for critical momentum $k_c$.} 
	\label{condition}	
\end{figure*}

 We now clarify that, to filter out the accidental DQPTs, the slow quench rate does not have to be infinitely small to make the quench dynamics adiabatic. As shown in Fig.~\ref{condition}, we present three-dimensional plots of the quantities \(|p_{k,i}||c_{k,i}|\) and the occupation amplitudes \(|p_{k,i}|^2\) for quench rates \(\beta = 0.08\) (upper panels a1–a2) and \(\beta = 0.1\) (bottom panels b1–b2). In the vicinity of the threshold quench rate \(\beta_{\text{th}} = 0.09\), condition (\ref{cond}) for DQPTs ceases to be valid. This threshold \(\beta_{\text{th}} = 0.09\) characterizes a “sufficiently slow quench.” For quench rates above \(\beta_{\text{th}}\) (b1), condition (\ref{cond}) is satisfied, ensuring the occurrence of DQPTs. Conversely, below \(\beta_{\text{th}}\) (a1), condition (\ref{cond}) is violated, and no DQPTs appear. Meanwhile, we note that in (a2), the system does not enter the adiabatic regime, as the excitation amplitude \(|p_{k,2}|^2\) (blue region) remains finite.

\begin{figure}[htbp]
	\includegraphics[width=1\linewidth]{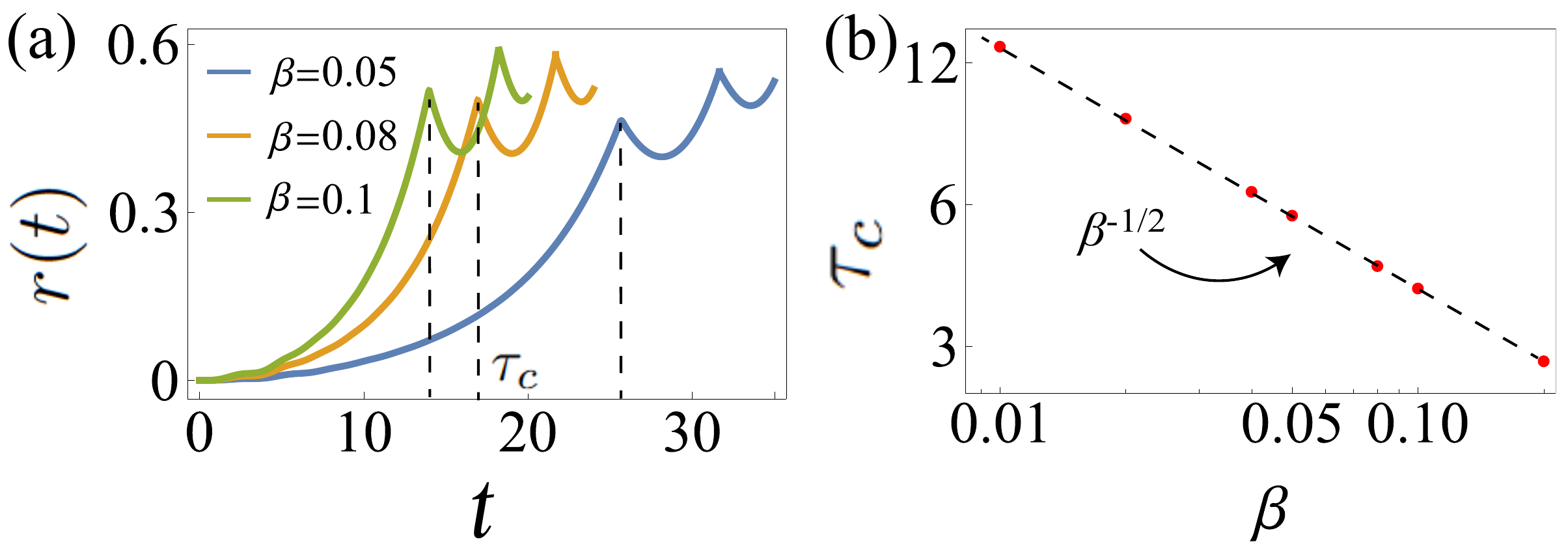}
	\caption{The scaling behavior of critical time of DQPT when the system is slowly quenched across the phase transition in the XY chain. (a) Rate function $r(t)$ calculated by the initial state $|\psi(t_i)\rangle$ with different quench rate. (b) The first critical time $\tau_c$ as a function of the quench rate $\beta$.}	
	\label{xy2}	
\end{figure}

Next, we further study the intrinsic connection between DQPT and equilibrium quantum phase transitions under the slow quench protocol and illustrate that the critical time $\tau_c$ of DQPTs exhibits a universal scaling law with respect to the quench rate $\beta$ \cite{zhang2025uni}. Remarkably, the exponent of this scaling law is governed by the critical exponent of the underlying equilibrium phase transitions, as shown in Fig.~\ref{xy2}. The first critical time $\tau_c$ exhibits a universal scaling with respect to the quench rate $\beta$, $\tau_c\sim \beta^{-\sigma}$ with $\sigma=1/2$. This result is consistent with the general argument \cite{zhang2025uni} that $\sigma=z\nu/(1+z\nu)$ with $z\nu=1$ in the XY chain model. Here, $z$ and $\nu$ are the critical exponents  of the equilibrium quantum phase transition.

\subsection{Aubry-André model}

We now investigate the quasiperiodic Aubry-André (AA) model, which is a toy model for the understanding of localization-delocalization transition. It is a chain of single-particle fermions with nearest-neighbor hopping $J$ subjected to on-site quasiperiodic potential $V_n$. The Hamiltonian is given by
\begin{equation}
    H=-J\sum_{n=1}^{N}\left(c_n^\dagger c_{n+1}+c_{n+1}^\dagger c_{n}\right)+\sum_{n=1}^{N}V_nc_n^\dagger c_{n},
\end{equation}
where $c_n^\dagger$ and $c_n$ are the free fermionic creation and annihilation operators at lattice site $n$, respectively, and $N$ is the lattice size. For simplicity, all energy scales are measured in units of $J$, which hereafter is set to unity. The on-site energy $V_n$ of the AA model is a quasiperiodic potential given by
\begin{equation}
    V_n=V\cos{(2\pi\alpha n+\phi)},
\end{equation}
where $V$ is the strength of the incommensurate potential, $\alpha=(\sqrt{5}-1)/2$ being an irrational number which is a deterministic quasiperiodic one, and $\phi$ is the phase parameter, which is set to zero without loss of any generality. The incommensurate potential can be viewed as a kind of quasi-random disorder, which drives the system undergoing a delocalization-localization transition at $V=2J$. When $V<2J$, all the eigenstates are extended, whereas all the eigenstates are localized, when $V>2J$. 

It has been demonstrated in Ref.~\cite{Yang2017dynamical} that  DQPTs can occur in the Aubry–André model when the system is suddenly quenched across the delocalization–localization phase boundary. Recently, Ref.~\cite{Ye2025disentangling} reported the emergence of energy-dependent DQPTs after sudden quenches that begin and end within the same extended phase. Here, the term ``energy-dependent" signifies that the manifestation of DQPTs is sensitive to the choice of the initial state. 

\begin{figure}[htbp]
	\includegraphics[width=1\linewidth]{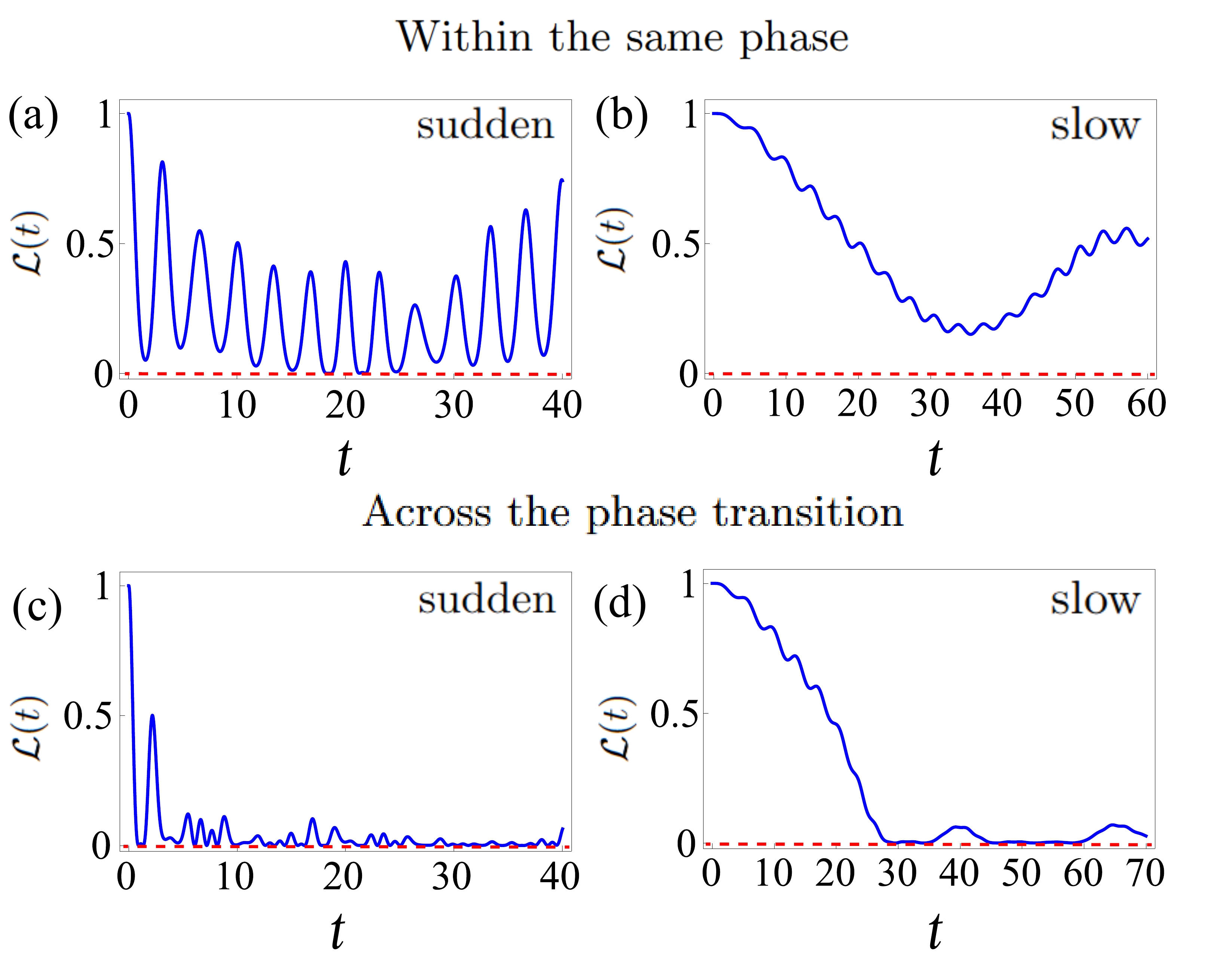}
	\caption{Loschmidt echo $\mathcal{L}(t)$ (blue lines) with sudden ((a) and (c)) and slow quench ((b) and (d)) in the Aubry-André model. (a) The commensurate potential $V$ is suddenly changed from $V_i=0$ to $V_f=1.8$ at time $t=0$. (b) Slow quench protocol: $V(t)=V_i+\beta t$, quench rate $\beta=0.1$. The pre-quench and post-quench Hamtonians are both are in the delocalization phase. (c) and (d) Quench protocol crosses the delocalization–localization phase boundary ($V_i=0 \to V_f=3$). Here, the red dashed lines represent $\mathcal{L}(t)=0$.}	
	\label{aah}	
\end{figure}

Here, we make a comprehensive study of DQPT in the AA model under both sudden and slow quenches of the potential strength $V$. We first consider the quench dynamics within the same delocalized phase. For sudden quench protocol within the same phase, one can observe that DQPTs occur at some critical times, while for slow quench protocols, DQPTs disappear, as shown in Fig.~\ref{aah}.   
 For the sudden quench, see Fig.~\ref{aah}(a), the Loschmidt echo $\mathcal{L}(t)$ becomes zero at some times (touching with the red dashed line), signaling energy-dependent DQPTs. In contrast, the Loschmidt echo $\mathcal{L}(t)$ remains finite at all times and can not touch with the zero axis under slow quench protocol, indicating that such energy-dependent DQPTs are filtered out when the quench rate is sufficiently slow, see Fig.~\ref{aah}(b). Here, we choose the initial state  to be the eigenstate corresponding to the middle eigenvalue of the Hamiltonian with system size $N=400$. 

We then study DQPT when the system is driven across the the delocalization–localization phase boundary.  Here the potential strength $V$ is either suddenly or slowly  quenched from $V_i=0$ in the delocalized phase to $V_f=3$ in the localized regime.  Fig.~\ref{aah}(c) displays the time evolution of the Loschmidt echo $\mathcal{L}(t)$ following a sudden quench. The recurrence of $\mathcal{L}(t)=0$ at multiple times signals the occurrence of DQPTs. However, some of these may be identified as accidental, as they are eliminated under  slow quench protocol [Fig.~\ref{aah}(d)]. Only those DQPTs, intrinsically linked to the delocalization–localization transition, remain robust under slow quench protocols.

\subsection{Trimer Su-Schrieffer-Heeger model}
In this subsection, we extend the discussion of DQPT from  two band models to multiband models by considering a three-band model, trimer Su-Schrieffer-Heeger model (SSH3) \cite{Jin2017top,Martinez2019edge,Anastasiadis2022bulk}. SSH3 is an extended SSH model with a unit cell that consists of three sites. The hopping between the sites is controlled by the couplings $t_1$ and $t_2$, while the different unit cells are coupled with the intercell coupling $t_3$. The system is governed by the Hamiltonian
\begin{equation}
	\begin{split}
		H=\sum_{j=1}^{N} \left(t_1 c_{j,B}^\dagger c_{j,A}+t_2 c_{j,C}^\dagger c_{j,B}+t_3 c_{j+1,A}^\dagger c_{j,C}\right)+{\rm H.c.}
	\end{split}
    \label{ssh3}
\end{equation}
Here, $c_{j,\alpha}^\dagger$ ($c_{j,\alpha}$) denotes the creation (annihilation) operator at site $\alpha$ (which can be either A, B or C type) of the $j$th unit cell. Without loss of generality, we will henceforth assume that the hopping parameters are real and non-negative. Considering periodic boundary conditions (PBC) along the length of the chain and performing Fourier transform of creation and annihilation operators, $c_{j,\alpha}^\dagger=(1/\sqrt{N})\sum_k e^{-ikj}c_{k,\alpha}^\dagger$, we can write the Hamiltonian in momentum space as $H=\sum_k\psi_k^\dagger H_k\psi_k$ with $\psi_k^\dagger=(c_{k,a}^\dagger,c_{k,b}^\dagger,c_{k,c}^\dagger)$, where
\begin{equation}
     H_k=\begin{pmatrix}
        0 & t_1 & t_3 e^{-ik} \\
        t_1 & 0 & t_2\\
        t_3 e^{ik} & t_2 & 0
        \end{pmatrix}.
\end{equation}

\begin{figure}[htbp]
	\includegraphics[width=1\linewidth]{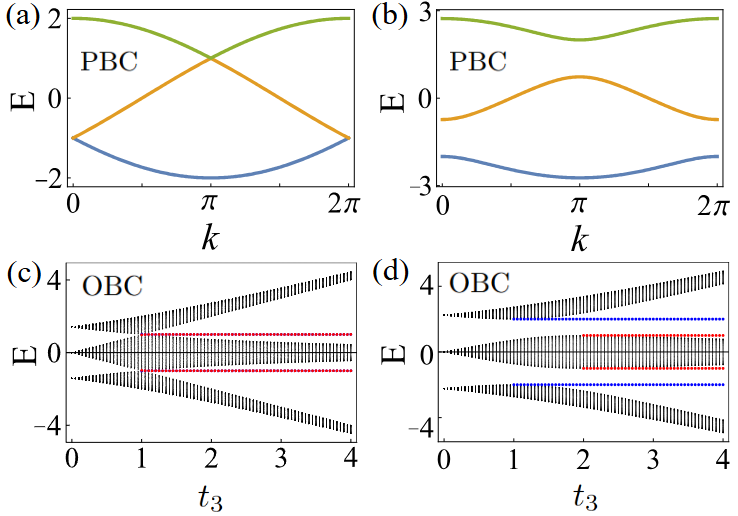}
	\caption{The spectrum of the bulk Hamiltonian of SSH3 in the first Brillouin zone under periodic boundary conditions for $t_1=t_2$ (a) and $t_1\neq t_2$ (b). (c) and (d) The real space energy spectrum with respect to the intercell hopping coupling $t_3$ for $t_1=t_2$ and $t_1\neq t_2$. Here, $N$ = 50. The red and blue dotted lines represent the topological edge states. }	
	\label{ssh3-energy}	
\end{figure}

For PBC, the spectrum of the above Hamiltonian consists of three dispersive bands are shown in Fig.~\ref{ssh3-energy}(a-b). The three bands touch each other at  $k=0$ and $\pi$, only when $t_1=t_2=t_3$, i.e., in the absence of trimerization. As the parameters are shifted from that condition ($t_1\neq t_2$), two band gaps appear in the band structure, as shown in Fig.~\ref{ssh3-energy}(b). Interestingly, these band gaps may host very peculiar edge states as those shown in Fig.~\ref{ssh3-energy}(d), where we have plotted the energy spectrum with open boundary conditions (OBC) of a finite trimer lattice ($N=50$). 

For OBC, we show the energy spectrum of a finite trimer lattice with $N = 50$ unit cells in Fig.~\ref{ssh3-energy}(c) and (d), as a function of the intercell hopping amplitude $t_3$ for two different cases with $t_1=t_2$ and $t_1\neq t_2$, respectively. We  observe a different number of in-gap edge states emerging as $t_3$ changes. In Fig.~\ref{ssh3-energy}(c), edge states (red dotted lines) can appear in the case of a mirror-symmetric SSH3 (i.e., when intracell coupling $t_1=t_2$) and they emerge at the point where the gap closes.  However, edge states can also appear in the case of an SSH3 (red dotted lines and blue dotted lines) that does not possess mirror symmetry (when intracell coupling $t_1\neq t_2$) and their emergence is not related to a gap closing (Fig.~\ref{ssh3-energy}(d)). 
In this case ($t_1\neq t_2$), the edge states appear after the passing of mirror-symmetric points, i.e., when $t_3=t_1$ and $t_3=t_2$.  Note that two-edge states (blue dotted lines) are localized on the right boundary of the system when intercell hopping $t_3$ satisfies $t_1<t_3<t_2$, and two new edge states (red dotted lines) can appear at the left boundary of the system for when $t_1<t_2<t_3$. The bulk-edge correspondence that can predict the emergence of all these edge states has been made via the sublattice Zak phase, which also remains valid in the absence of mirror symmetry \cite{Anastasiadis2022bulk}. 

\begin{figure}[htbp]
	\includegraphics[width=1\linewidth]{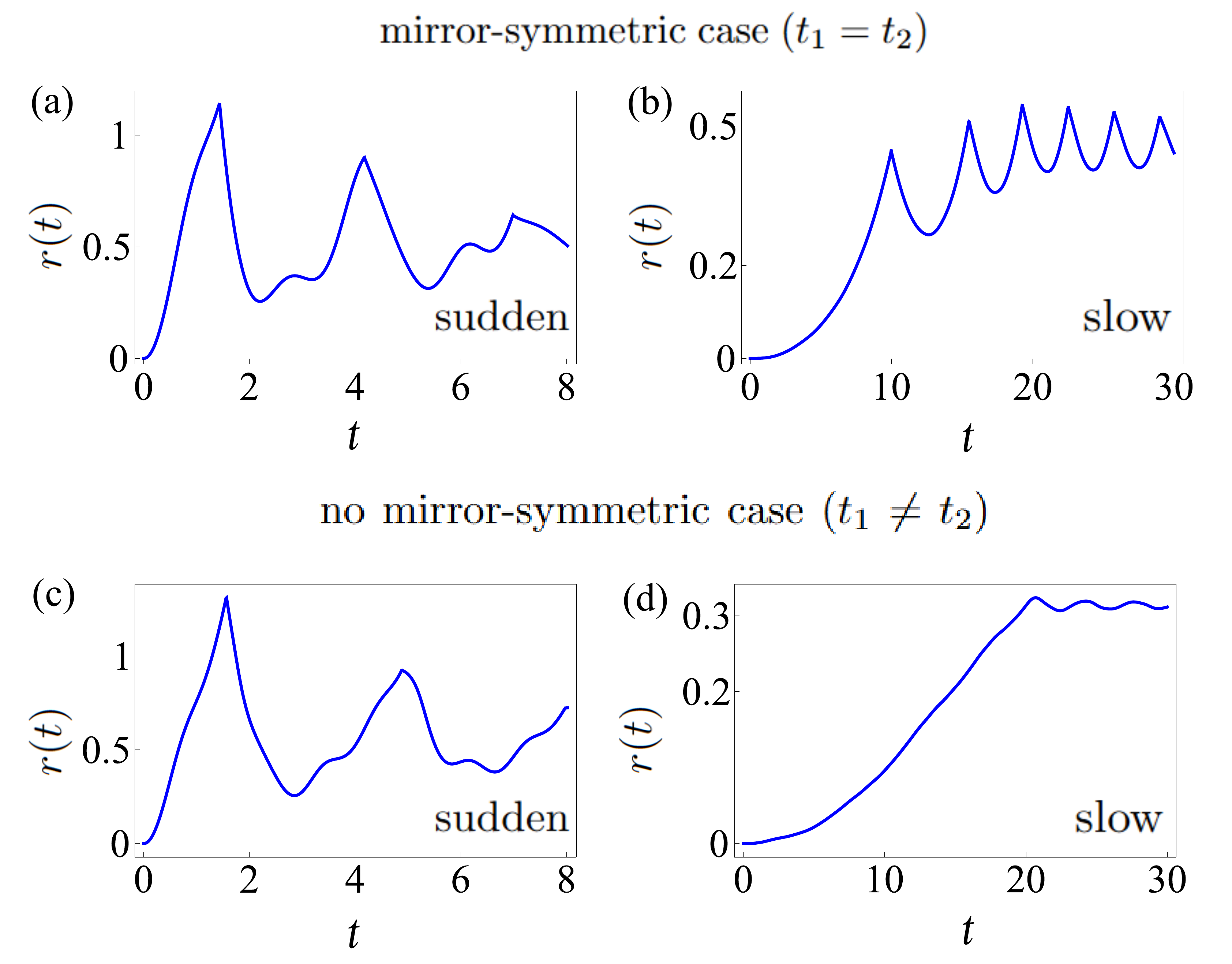}
	\caption{Sudden and slow quench dynamics in the SSH3 model for mirror-symmetric case $(t_1=t_2)$ (a-b) and no mirror-symmetric case $(t_1\neq t_2)$ (c-d).
    (a) The rate function $r(t)$ of the Loschmidt amplitude under sudden quench protocol. $t_3$ is suddenly changed from 0.5 to 2.5 at time $t=0$. (b) The rate function $r(t)$ of the Loschmidt amplitude under slow quench protocol. $t_3(t)=t_{3,i}+\beta t$ with quench rate $\beta=0.1$. (c) The sudden quench protocol is same as (a). (d) The slow quench protocol is also same as (b) with quench rate $\beta=0.1$.}	
	\label{ssh3-symmetry}	
\end{figure}


We now examine both sudden and slow quench protocols for the mirror-symmetric case ($t_1= t_2$). The parameter $t_3$ is quenched across the topological phase boundary, a process accompanied by the closure of the energy gap at the critical point $t_3=t_1=t_2$. Fig.~\ref{ssh3-symmetry} (a) depictes the rate function $r(t)$ for the sudden quench protocol, while (b) presentes the corresponding results for the slow quench protocol.
In this mirror-symmetric case, DQPT is protected by the symmetry of the Hamiltonian and the critical times of DQPTs present a distinct periodicity. 


We next investigate sudden and slow quench protocols in the absence of mirror symmetry ($t_1\neq t_2$). Although edge states may emerge in real space, their appearance is not associated with a bulk gap closing. This leads to markedly distinct dynamical behaviors between the two quench protocols. As shown in Figs.~\ref{ssh3-symmetry}(c), DQPTs emerge under sudden quench. In contrast, DQPTs are not observed under slow quench, as evidenced in Figs.~\ref{ssh3-symmetry}(d). We emphasize that the connection between slow-quench DQPT and equilibrium phase transitions fundamentally relies on the occurrence of a bulk gap closure; with its absence, such dynamical signatures do not faithfully reflect the underlying equilibrium criticality.



\begin{figure}[htbp]
	\includegraphics[width=1\linewidth]{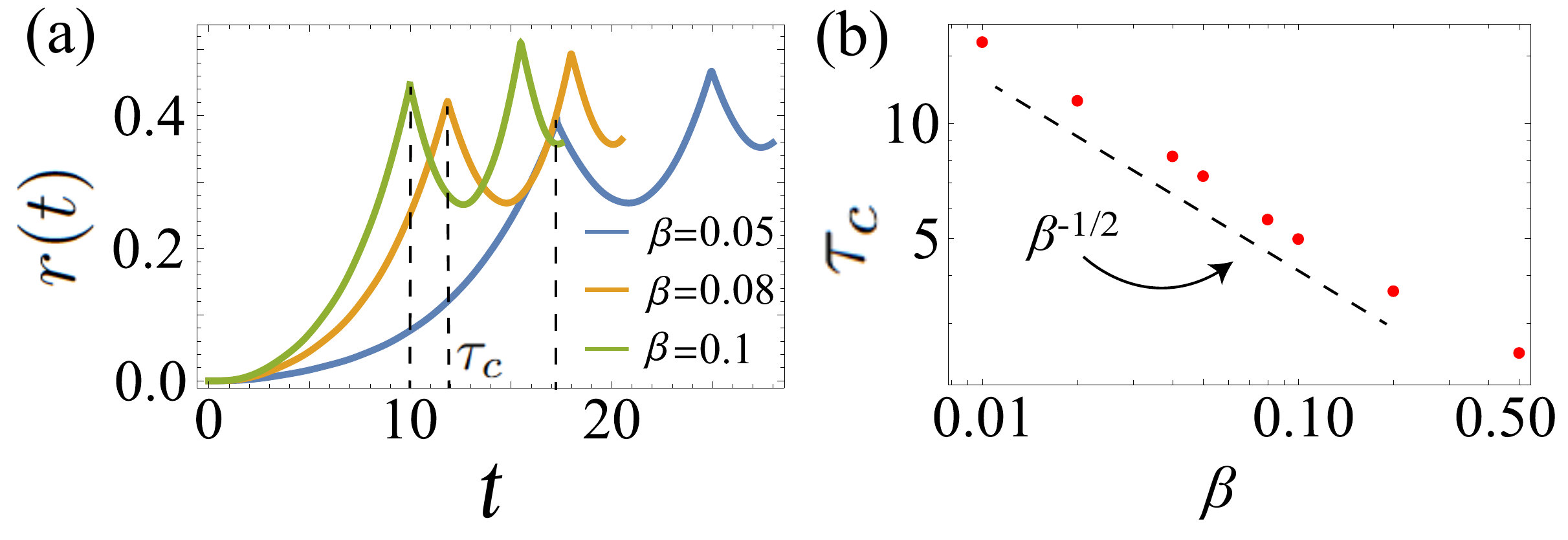}
	\caption{ Rate function $r(t)$ (a) and the scaling behavior of the critical time $\tau_c$ (b) of Loschmidt amplitude $\mathcal{L}(t)$ in the trimer Su-Schrieffer-Heeger model in the mirror-symmetric case ($t_1 = t_2$) under slow quench protocol. (a) Rate function $r(t)$ for different quench rates $\beta$. The vertical black dashed lines correspond to the first critical times $\tau_c$.}	
	\label{ssh3-scaling}	
\end{figure}

Beyond the previously discussed advantages of slow quench protocols, the critical time $\tau_c$ of dynamical quantum phase transitions can be leveraged to extract the critical exponent of the corresponding equilibrium quantum phase transition \cite{zhang2025uni}. We pay our attention to the mirror-symmetric regime and use the ground state of the pre-quench Hamiltonian as the initial wavefunction. The rate function $r(t)$ with different quench rate $\beta$ is shown in Fig.~\ref{ssh3-scaling}(a). By extracting the first critical time
we observe a universal scaling relation:
\begin{equation}
   \tau_c\sim\beta^{-1/2}+\tau_0.
\end{equation}
as illustrated in Fig.~\ref{ssh3-scaling} (b). Here, $\tau_0$ is an constant shift. This scaling behavior yields the critical exponent product $z\nu=1$, consistent with the universality class reported in \cite{zhang2025uni}. This observation further indicates that this connection between DQPT and equilibrium phase transitions is not only valid in two-band models but also in multiband models.

\section{Conclusion}
\label{s4}
In summary, we investigate the fate of dynamical quantum phase transitions when the quench protocol is switched from sudden to slow quench limit. We have made a general argument that accidental DQPTs can be filtered out under the slow quench protocol. We have investigated three distinct models to illustrate our finding: the XY chain, the Aubry-André model, and the trimer Su-Schrieffer-Heeger model by performing a comprehensive numerical study of DQPT under both sudden and slow quench protocols. For the XY chain \cite{Vajna2014dis},   previous studies  have reported that DQPT can occur under sudden quench when the pre-quench and post-quench Hamiltonian belong to the same ferromagnetic phase. 
For the Aubry-André model, Ref.~\cite{Ye2025disentangling} reported the emergence of energy-dependent dynamical quantum phase transitions during sudden quenches that begin and end within the same extended phase. Under slow quench dynamics, we show that both above cases of DQPTs are accidental and can be filtered out.
  For the trimer Su-Schrieffer-Heeger model, there exist two different kinds of  topological phase transitions. One is accompanied by a bulk gap closure and the other one is not. We emphasize that the connection between slow-quench DQPT and equilibrium phase transitions fundamentally relies on the occurrence of a bulk gap closure; in its absence, such dynamical signatures do not faithfully reflect the underlying equilibrium criticality. Therefore, we have  demonstrated that slow quenche protocol can be used to reveal the intrinsic connection between DQPT and equilibrium quantum phase transitions, thereby enabling DQPT to serve as a dynamical probe for identifying the equilibrium quantum phase transition and the corresponding critical exponents.

\section*{Acknowledgements}
This work was supported by the National Natural Science Foundation of China (No. 12275075), the National Key Research and Development Program of Ministry of Science and Technology (No. 2021YFA1200700), the China Postdoctoral Science Foundation (Grant No. 2021M690970) and the Fundamental Research Funds for the Central Universities from China. \\

\section*{DATAAVAILABILITY}
All the data that support the plots and other findings of this study are available from the corresponding author upon reasonable request.

\bibliography{ref}

@article{Langen,
  title={Ultracold atoms out of equilibrium},
  author={Langen, Tim and Geiger, Remi and Schmiedmayer, J{\"o}rg},
  journal={Annu. Rev. Condens. Matter Phys.},
  volume={6},
  number={1},
  pages={201--217},
  year={2015},
  publisher={Annual Reviews}
}

@article{Leibfried,
  title = {Quantum dynamics of single trapped ions},
  author = {Leibfried, D. and Blatt, R. and Monroe, C. and Wineland, D.},
  journal = {Rev. Mod. Phys.},
  volume = {75},
  issue = {1},
  pages = {281--324},
  numpages = {0},
  year = {2003},
  month = {Mar},
  publisher = {American Physical Society},
}

@article{Liu2024,
  title = {{Dynamical Transition Due} to {Feedback-Induced Skin Effect}},
  author = {Liu, Ze-Chuan and Li, Kai and Xu, Yong},
  journal = {Phys. Rev. Lett.},
  volume = {133},
  issue = {9},
  pages = {090401},
  numpages = {6},
  year = {2024},
  month = {Aug},
  publisher = {American Physical Society},
}

@article{Zache2019,
  title = {Dynamical {Topological Transitions} in the {Massive Schwinger Model} with a $\ensuremath{\theta}$ Term},
  author = {Zache, T. V. and Mueller, N. and Schneider, J. T. and Jendrzejewski, F. and Berges, J. and Hauke, P.},
  journal = {Phys. Rev. Lett.},
  volume = {122},
  issue = {5},
  pages = {050403},
  numpages = {6},
  year = {2019},
  month = {Feb},
  publisher = {American Physical Society},
}

@article{Pal,
  title = {Many-body localization phase transition},
  author = {Pal, Arijeet and Huse, David A.},
  journal = {Phys. Rev. B},
  volume = {82},
  issue = {17},
  pages = {174411},
  numpages = {7},
  year = {2010},
  month = {Nov},
  publisher = {American Physical Society},
}

@article{Nandkishore,
  title={Many-body localization and thermalization in quantum statistical mechanics},
  author={Nandkishore, Rahul and Huse, David A},
  journal={Annu. Rev. Condens. Matter Phys.},
  volume={6},
  number={1},
  pages={15--38},
  year={2015},
  publisher={Annual Reviews}
}

@article{Abanin,
  title = {Colloquium: {Many-body} localization, thermalization, and entanglement},
  author = {Abanin, Dmitry A. and Altman, Ehud and Bloch, Immanuel and Serbyn, Maksym},
  journal = {Rev. Mod. Phys.},
  volume = {91},
  issue = {2},
  pages = {021001},
  numpages = {26},
  year = {2019},
  month = {May},
  publisher = {American Physical Society},
}

@article{Schreiber,
  title={Observation of many-body localization of interacting fermions in a quasirandom optical lattice},
  author={Schreiber, Michael and Hodgman, Sean S and Bordia, Pranjal and L{\"u}schen, Henrik P and Fischer, Mark H and Vosk, Ronen and Altman, Ehud and Schneider, Ulrich and Bloch, Immanuel},
  journal={Science},
  volume={349},
  number={6250},
  pages={842--845},
  year={2015},
  publisher={American Association for the Advancement of Science}
}

@article{Smith,
  title={Many-body localization in a quantum simulator with programmable random disorder},
  author={Smith, Jacob and Lee, Aaron and Richerme, Philip and Neyenhuis, Brian and Hess, Paul W and Hauke, Philipp and Heyl, Markus and Huse, David A and Monroe, Christopher},
  journal={Nature Physics},
  volume={12},
  number={10},
  pages={907--911},
  year={2016},
  publisher={Nature Publishing Group UK London}
}

@article{Choi,
  title={Exploring the many-body localization transition in two dimensions},
  author={Choi, Jae-yoon and Hild, Sebastian and Zeiher, Johannes and Schau{\ss}, Peter and Rubio-Abadal, Antonio and Yefsah, Tarik and Khemani, Vedika and Huse, David A and Bloch, Immanuel and Gross, Christian},
  journal={Science},
  volume={352},
  number={6293},
  pages={1547--1552},
  year={2016},
  publisher={American Association for the Advancement of Science}
}

@article{Zurek2005,
  title={Dynamics of a quantum phase transition},
  author={Zurek, Wojciech H and Dorner, Uwe and Zoller, Peter},
  journal={Phys. Rev. Lett.},
  volume={95},
  number={10},
  pages={105701},
  year={2005},
  publisher={APS}
}

@article{Polkovnikov2005,
  title={Universal adiabatic dynamics in the vicinity of a quantum critical point},
  author={Polkovnikov, Anatoli},
  journal={Phys. Rev. B},
  volume={72},
  number={16},
  pages={161201},
  year={2005},
  publisher={APS}
}

@article{Sen2008,
  title={Defect production in nonlinear quench across a quantum critical point},
  author={Sen, Diptiman and Sengupta, K and Mondal, Shreyoshi},
  journal={Phys. Rev. Lett.},
  volume={101},
  number={1},
  pages={016806},
  year={2008},
  publisher={APS}
}

@article{Barankov2008,
  title={Optimal nonlinear passage through a quantum critical point},
  author={Barankov, Roman and Polkovnikov, Anatoli},
  journal={Phys. Rev. Lett.},
  volume={101},
  number={7},
  pages={076801},
  year={2008},
  publisher={APS}
}

@article{Dziarmaga2010,
  title={Dynamics of a quantum phase transition and relaxation to a steady state},
  author={Dziarmaga, Jacek},
  journal={Advances in Physics},
  volume={59},
  number={6},
  pages={1063--1189},
  year={2010},
  publisher={Taylor \& Francis}
}

@article{Polkovnikov2011,
  title={Colloquium: {Nonequilibrium} dynamics of closed interacting quantum systems},
  author={Polkovnikov, Anatoli and Sengupta, Krishnendu and Silva, Alessandro and Vengalattore, Mukund},
  journal={Rev. Mod. Phys.},
  volume={83},
  number={3},
  pages={863--883},
  year={2011},
  publisher={APS}
}

@article{Nowak2021,
  title={Quantum {Kibble-Zurek} mechanism: {Kink} correlations after a quench in the quantum Ising chain},
  author={Nowak, Rados{\l}aw J and Dziarmaga, Jacek},
  journal={Phys. Rev. B},
  volume={104},
  number={7},
  pages={075448},
  year={2021},
  publisher={APS}
}

@article{Kou2023,
  title={Varying quench dynamics in the transverse {Ising} chain: {The Kibble-Zurek}, saturated, and presaturated regimes},
  author={Kou, Han-Chuan and Li, Peng},
  journal={Phys. Rev. B},
  volume={108},
  number={21},
  pages={214307},
  year={2023},
  publisher={APS}
}

@article{Liang2024,
  title={Quantum criticality and {Kibble-Zurek} scaling in the {Aubry-Andr{\'e}-Stark} model},
  author={Liang, En-Wen and Tang, Ling-Zhi and Zhang, Dan-Wei},
  journal={Phys. Rev. B},
  volume={110},
  number={2},
  pages={024207},
  year={2024},
  publisher={APS}
}

@article{Wilczek,
  title={Quantum time crystals},
  author={Wilczek, Frank},
  journal={Phys. Rev. Lett.},
  volume={109},
  number={16},
  pages={160401},
  year={2012},
  publisher={APS}
}

@book{Sacha,
  title={Time crystals},
  author={Sacha, Krzysztof},
  volume={114},
  year={2020},
  publisher={Springer}
}

@article{Zaletel,
  title={Colloquium: {Quantum} and classical discrete time crystals},
  author={Zaletel, Michael P and Lukin, Mikhail and Monroe, Christopher and Nayak, Chetan and Wilczek, Frank and Yao, Norman Y},
  journal={Rev. Mod. Phys.},
  volume={95},
  number={3},
  pages={031001},
  year={2023},
  publisher={APS}
}

@article{Heyl2018dynamical,
  title={Dynamical quantum phase transitions in spin chains with long-range interactions: {Merging} different concepts of nonequilibrium criticality},
  author={{\v{Z}}unkovi{\v{c}}, Bojan and Heyl, Markus and Knap, Michael and Silva, Alessandro},
  journal={Phys. Rev. Lett.},
  volume={120},
  number={13},
  pages={130601},
  year={2018},
  publisher={APS}
}

@article{Lang2018dynamical,
  title={Dynamical quantum phase transitions: {A} geometric picture},
  author={Lang, Johannes and Frank, Bernhard and Halimeh, Jad C},
  journal={Phys. Rev. Lett.},
  volume={121},
  number={13},
  pages={130603},
  year={2018},
  publisher={APS}
}

@article{Bandyopadhyay2021ob,
  title={Observing dynamical quantum phase transitions through quasilocal string operators},
  author={Bandyopadhyay, Souvik and Polkovnikov, Anatoli and Dutta, Amit},
  journal={Phys. Rev. Lett.},
  volume={126},
  number={20},
  pages={200602},
  year={2021},
  publisher={APS}
}

@article{Vajna2015topo,
  title={Topological classification of dynamical phase transitions},
  author={Vajna, Szabolcs and D{\'o}ra, Bal{\'a}zs},
  journal={Phys. Rev. B},
  volume={91},
  number={15},
  pages={155127},
  year={2015},
  publisher={APS}
}

@article{Vosk2014dynamical,
  title = {Dynamical {Quantum Phase Transitions} in {Random Spin Chains}},
  author = {Vosk, Ronen and Altman, Ehud},
  journal = {Phys. Rev. Lett.},
  volume = {112},
  issue = {21},
  pages = {217204},
  numpages = {5},
  year = {2014},
  month = {May},
  publisher = {American Physical Society}
}

@article{Bhattacharya2017inter,
  title={Interconnections between equilibrium topology and dynamical quantum phase transitions in a linearly ramped {Haldane} model},
  author={Bhattacharya, Utso and Dutta, Amit},
  journal={Phys. Rev. B},
  volume={95},
  number={18},
  pages={184307},
  year={2017},
  publisher={APS}
}

@article{Vajna2014dis,
  title={Disentangling dynamical phase transitions from equilibrium phase transitions},
  author={Vajna, Szabolcs and D{\'o}ra, Bal{\'a}zs},
  journal={Phys. Rev. B},
  volume={89},
  number={16},
  pages={161105},
  year={2014},
  publisher={APS}
}

@article{Huang2019dynamical,
  title={Dynamical quantum phase transitions in {U} (1) quantum link models},
  author={Huang, Yi-Ping and Banerjee, Debasish and Heyl, Markus},
  journal={Phys. Rev. Lett.},
  volume={122},
  number={25},
  pages={250401},
  year={2019},
  publisher={APS}
}

@article{Link2020dynamical,
  title={Dynamical phase transitions in dissipative quantum dynamics with quantum optical realization},
  author={Link, Valentin and Strunz, Walter T},
  journal={Phys. Rev. Lett.},
  volume={125},
  number={14},
  pages={143602},
  year={2020},
  publisher={APS}
}

@article{Heyl2018dyn,
  title={Dynamical quantum phase transitions: a review},
  author={Heyl, Markus},
  journal={Reports on Progress in Physics},
  volume={81},
  number={5},
  pages={054001},
  year={2018},
  publisher={IOP Publishing}
}

@article{Jurcevic2017direct,
  title={Direct observation of dynamical quantum phase transitions in an interacting many-body system},
  author={Jurcevic, P and Shen, H and Hauke, P and Maier, C and Brydges, T and Hempel, C and Lanyon, BP and Heyl, Markus and Blatt, R and Roos, CF},
  journal={Phys. Rev. Lett.},
  volume={119},
  number={8},
  pages={080501},
  year={2017},
  publisher={APS}
}

@article{Flaschner2018observation,
  title={Observation of dynamical vortices after quenches in a system with topology},
  author={Fl{\"a}schner, N and Vogel, D and Tarnowski, M and Rem, BS and L{\"u}hmann, D-S and Heyl, Markus and Budich, JC and Mathey, L and Sengstock, Klaus and Weitenberg, C},
  journal={Nature Physics},
  volume={14},
  number={3},
  pages={265--268},
  year={2018},
  publisher={Nature Publishing Group UK London}
}

@article{Wang2019simulating,
  title={Simulating dynamic quantum phase transitions in photonic quantum walks},
  author={Wang, Kunkun and Qiu, Xingze and Xiao, Lei and Zhan, Xiang and Bian, Zhihao and Yi, Wei and Xue, Peng},
  journal={Phys. Rev. Lett.},
  volume={122},
  number={2},
  pages={020501},
  year={2019},
  publisher={APS}
}

@article{Huang2016dynamical,
  title={Dynamical quantum phase transitions: role of topological nodes in wave function overlaps},
  author={Huang, Zhoushen and Balatsky, Alexander V},
  journal={Phys. Rev. Lett.},
  volume={117},
  number={8},
  pages={086802},
  year={2016},
  publisher={APS}
}

@article{Homrighausen2017,
  title={Anomalous dynamical phase in quantum spin chains with long-range interactions},
  author={Homrighausen, Ingo and Abeling, Nils O and Zauner-Stauber, Valentin and Halimeh, Jad C},
  journal={Phys. Rev. B},
  volume={96},
  number={10},
  pages={104436},
  year={2017},
  publisher={APS}
}

@article{Nicola2021entanglement,
  title={Entanglement view of dynamical quantum phase transitions},
  author={De Nicola, Stefano and Michailidis, Alexios A and Serbyn, Maksym},
  journal={Phys. Rev. Lett.},
  volume={126},
  number={4},
  pages={040602},
  year={2021},
  publisher={APS}
}

@article{Heyl2013,
  title={Dynamical quantum phase transitions in the transverse-field {Ising} model},
  author={Heyl, Markus and Polkovnikov, Anatoli and Kehrein, Stefan},
  journal={Phys. Rev. Lett.},
  volume={110},
  number={13},
  pages={135704},
  year={2013},
  publisher={APS}
}

@article{Sharma2016,
  title={Slow quenches in a quantum {Ising} chain: {Dynamical} phase transitions and topology},
  author={Sharma, Shraddha and Divakaran, Uma and Polkovnikov, Anatoli and Dutta, Amit},
  journal={Phys. Rev. B},
  volume={93},
  number={14},
  pages={144306},
  year={2016},
  publisher={APS}
}

@article{Puskarov2016,
  title={Time evolution during and after finite-time quantum quenches in the transverse-field {Ising} chain},
  author={Puskarov, Tatjana and Schuricht, Dirk},
  journal={SciPost Physics},
  volume={1},
  number={1},
  pages={003},
  year={2016}
}

@article{Gorin2006dynamics,
  title={Dynamics of {Loschmidt} echoes and fidelity decay},
  author={Gorin, Thomas and Prosen, Toma{\v{z}} and Seligman, Thomas H and {\v{Z}}nidari{\v{c}}, Marko},
  journal={Physics Reports},
  volume={435},
  number={2-5},
  pages={33--156},
  year={2006},
  publisher={Elsevier}
}

@article{Quan2006decay,
  title = {Decay of {Loschmidt Echo Enhanced} by {Quantum Criticality}},
  author = {Quan, H. T. and Song, Z. and Liu, X. F. and Zanardi, P. and Sun, C. P.},
  journal = {Phys. Rev. Lett.},
  volume = {96},
  issue = {14},
  pages = {140604},
  numpages = {4},
  year = {2006},
  month = {Apr},
  publisher = {American Physical Society}
}

@article{Hasegawa2021irreversibility,
  title = {Irreversibility, {Loschmidt Echo}, and {Thermodynamic Uncertainty Relation}},
  author = {Hasegawa, Yoshihiko},
  journal = {Phys. Rev. Lett.},
  volume = {127},
  issue = {24},
  pages = {240602},
  numpages = {7},
  year = {2021},
  month = {Dec},
  publisher = {American Physical Society},
}

@article{Jafari2017loschmidt,
  title = {Loschmidt {Echo Revivals}: {Critical} and {Noncritical}},
  author = {Jafari, R. and Johannesson, Henrik},
  journal = {Phys. Rev. Lett.},
  volume = {118},
  issue = {1},
  pages = {015701},
  numpages = {5},
  year = {2017},
  month = {Jan},
  publisher = {American Physical Society}
}

@article{Zurek2001sub,
  title={Sub-{Planck} structure in phase space and its relevance for quantum decoherence},
  author={Zurek, Wojciech Hubert},
  journal={Nature},
  volume={412},
  number={6848},
  pages={712--717},
  year={2001},
  publisher={Nature Publishing Group UK London}
}

@article{Zurek2003decoherence,
  title = {Decoherence, einselection, and the quantum origins of the classical},
  author = {Zurek, Wojciech Hubert},
  journal = {Rev. Mod. Phys.},
  volume = {75},
  issue = {3},
  pages = {715--775},
  numpages = {0},
  year = {2003},
  month = {May},
  publisher = {American Physical Society}
}

@article{Cucchietti2003decoherence,
  title = {Decoherence and the {Loschmidt Echo}},
  author = {Cucchietti, F. M. and Dalvit, D. A. R. and Paz, J. P. and Zurek, W. H.},
  journal = {Phys. Rev. Lett.},
  volume = {91},
  issue = {21},
  pages = {210403},
  numpages = {4},
  year = {2003},
  month = {Nov},
  publisher = {American Physical Society}
}

@article{Gutierrez2009long,
  title = {Long-time saturation of the {Loschmidt} echo in quantum chaotic billiards},
  author = {Guti\'errez, Martha and Goussev, Arseni},
  journal = {Phys. Rev. E},
  volume = {79},
  issue = {4},
  pages = {046211},
  numpages = {5},
  year = {2009},
  month = {Apr},
  publisher = {American Physical Society}
}

@article{Jacquod2001golden,
  title = {Golden rule decay versus {Lyapunov} decay of the quantum {Loschmidt echo}},
  author = {Jacquod, Ph. and Silvestrov, P.G. and Beenakker, C.W.J.},
  journal = {Phys. Rev. E},
  volume = {64},
  issue = {5},
  pages = {055203},
  numpages = {4},
  year = {2001},
  month = {Oct},
  publisher = {American Physical Society}
}

@article{Cerruti2002sensitivity,
  title = {Sensitivity of {Wave Field Evolution} and {Manifold Stability} in {Chaotic Systems}},
  author = {Cerruti, Nicholas R. and Tomsovic, Steven},
  journal = {Phys. Rev. Lett.},
  volume = {88},
  issue = {5},
  pages = {054103},
  numpages = {4},
  year = {2002},
  month = {Jan},
  publisher = {American Physical Society}
}

@article{Wisniacki2003short,
  title = {Short-time decay of the {Loschmidt} echo},
  author = {Wisniacki, Diego A.},
  journal = {Phys. Rev. E},
  volume = {67},
  issue = {1},
  pages = {016205},
  numpages = {6},
  year = {2003},
  month = {Jan},
  publisher = {American Physical Society}
}

@article{Cucchietti2004universality,
  title = {Universality of the {Lyapunov} regime for the {Loschmidt} echo},
  author = {Cucchietti, Fernando M. and Pastawski, Horacio M. and Jalabert, Rodolfo A.},
  journal = {Phys. Rev. B},
  volume = {70},
  issue = {3},
  pages = {035311},
  numpages = {23},
  year = {2004},
  month = {Jul},
  publisher = {American Physical Society}
}

@article{tang2022dynamical,
  title={{Dynamical} scaling of {Loschmidt echo} in non-Hermitian systems},
  author={Tang, Jia-Chen and Kou, Su-Peng and Sun, Gaoyong},
  journal={Europhysics Letters},
  volume={137},
  number={4},
  pages={40001},
  year={2022},
  publisher={IOP Publishing}
}

@article{Zener1932non,
  title={Non-adiabatic crossing of energy levels},
  author={Zener, Clarence},
  journal={Proc. R. Soc. A},
  volume={137},
  number={833},
  pages={696--702},
  year={1932},
  publisher={The Royal Society London}
}

@article{Majorana1932atomi,
  title={Atomi orientati in campo magnetico variabile},
  author={Majorana, Ettore},
  journal={Il Nuovo Cimento},
  volume={9},
  number={2},
  pages={43--50},
  year={1932},
  publisher={Springer}
}

@article{Dziarmaga2005,
  title = {Dynamics of a Quantum Phase Transition: Exact Solution of the Quantum Ising Model},
  author = {Dziarmaga, Jacek},
  journal = {Phys. Rev. Lett.},
  volume = {95},
  issue = {24},
  pages = {245701},
  numpages = {4},
  year = {2005},
  month = {Dec},
  publisher = {American Physical Society}
}

@article{Jin2017top,
  title = {Topological phases and edge states in a non-{Hermitian} trimerized optical lattice},
  author = {Jin, L.},
  journal = {Phys. Rev. A},
  volume = {96},
  issue = {3},
  pages = {032103},
  numpages = {8},
  year = {2017},
  month = {Sep},
  publisher = {American Physical Society}
}

@article{Martinez2019edge,
  title = {Edge states in trimer lattices},
  author = {Martinez Alvarez, V. M. and Coutinho-Filho, M. D.},
  journal = {Phys. Rev. A},
  volume = {99},
  issue = {1},
  pages = {013833},
  numpages = {8},
  year = {2019},
  month = {Jan},
  publisher = {American Physical Society}
}

@article{Anastasiadis2022bulk,
  title = {Bulk-edge correspondence in the trimer {Su-Schrieffer-Heeger} model},
  author = {Anastasiadis, Adamantios and Styliaris, Georgios and Chaunsali, Rajesh and Theocharis, Georgios and Diakonos, Fotios K.},
  journal = {Phys. Rev. B},
  volume = {106},
  issue = {8},
  pages = {085109},
  numpages = {14},
  year = {2022},
  month = {Aug},
  publisher = {American Physical Society}
}

@article{zhang2025uni,
  title = {Universal scaling in Loschmidt echo across quantum phase transitions under nonadiabatic dynamics},
  author = {Zhang, Xiang and Hu, Liangdong and Li, Fuxiang},
  journal = {Phys. Rev. B},
  volume = {112},
  issue = {2},
  pages = {024310},
  numpages = {14},
  year = {2025},
  month = {Jul},
  publisher = {American Physical Society}
}

@article{Nie2020experimental,
  title = {Experimental {Observation of Equilibrium and Dynamical Quantum Phase Transitions via Out-of-Time-Ordered Correlators}},
  author = {Nie, Xinfang and Wei, Bo-Bo and Chen, Xi and Zhang, Ze and Zhao, Xiuzhu and Qiu, Chudan and Tian, Yu and Ji, Yunlan and Xin, Tao and Lu, Dawei and Li, Jun},
  journal = {Phys. Rev. Lett.},
  volume = {124},
  issue = {25},
  pages = {250601},
  numpages = {6},
  year = {2020},
  month = {Jun},
  publisher = {American Physical Society}
}

@article{Guo20190bservation,
  title = {Observation of a {Dynamical Quantum Phase Transition by a Superconducting Qubit Simulation}},
  author = {Guo, Xue-Yi and Yang, Chao and Zeng, Yu and Peng, Yi and Li, He-Kang and Deng, Hui and Jin, Yi-Rong and Chen, Shu and Zheng, Dongning and Fan, Heng},
  journal = {Phys. Rev. Appl.},
  volume = {11},
  issue = {4},
  pages = {044080},
  numpages = {12},
  year = {2019},
  month = {Apr},
  publisher = {American Physical Society}
}

@article{xu2020measuring,
  title={Measuring a dynamical topological order parameter in quantum walks},
  author={Xu, Xiao-Ye and Wang, Qin-Qin and Heyl, Markus and Budich, Jan Carl and Pan, Wei-Wei and Chen, Zhe and Jan, Munsif and Sun, Kai and Xu, Jin-Shi and Han, Yong-Jian and others},
  journal={Light: Science \& Applications},
  volume={9},
  number={1},
  pages={7},
  year={2020},
  publisher={Nature Publishing Group UK London}
}

@article{Yang2019floquet,
  title = {Floquet dynamical quantum phase transitions},
  author = {Yang, Kai and Zhou, Longwen and Ma, Wenchao and Kong, Xi and Wang, Pengfei and Qin, Xi and Rong, Xing and Wang, Ya and Shi, Fazhan and Gong, Jiangbin and Du, Jiangfeng},
  journal = {Phys. Rev. B},
  volume = {100},
  issue = {8},
  pages = {085308},
  numpages = {11},
  year = {2019},
  month = {Aug},
  publisher = {American Physical Society}
}

@article{Zamani2020floquet,
  title = {Floquet dynamical quantum phase transition in the extended {XY} model: {Nonadiabatic} to adiabatic topological transition},
  author = {Zamani, Sara and Jafari, R. and Langari, A.},
  journal = {Phys. Rev. B},
  volume = {102},
  issue = {14},
  pages = {144306},
  numpages = {12},
  year = {2020},
  month = {Oct},
  publisher = {American Physical Society}
}

@article{Jafari2021floquet,
  title = {Floquet dynamical phase transition and entanglement spectrum},
  author = {Jafari, R. and Akbari, Alireza},
  journal = {Phys. Rev. A},
  volume = {103},
  issue = {1},
  pages = {012204},
  numpages = {8},
  year = {2021},
  month = {Jan},
  publisher = {American Physical Society}
}

@article{Zamani2022out,
  title = {Out-of-time-order correlations and {Floquet} dynamical quantum phase transition},
  author = {Zamani, Sara and Jafari, R. and Langari, A.},
  journal = {Phys. Rev. B},
  volume = {105},
  issue = {9},
  pages = {094304},
  numpages = {15},
  year = {2022},
  month = {Mar},
  publisher = {American Physical Society}
}

@article{Jafari2022floquet,
  title = {Floquet dynamical quantum phase transitions under synchronized periodic driving},
  author = {Jafari, R. and Akbari, Alireza and Mishra, Utkarsh and Johannesson, Henrik},
  journal = {Phys. Rev. B},
  volume = {105},
  issue = {9},
  pages = {094311},
  numpages = {13},
  year = {2022},
  month = {Mar},
  publisher = {American Physical Society}
}

@article{Mera2018dynamical,
  title = {Dynamical phase transitions at finite temperature from fidelity and interferometric Loschmidt echo induced metrics},
  author = {Mera, Bruno and Vlachou, Chrysoula and Paunkovi\ifmmode \acute{c}\else \'{c}\fi{}, Nikola and Vieira, V\'{\i}tor R. and Viyuela, Oscar},
  journal = {Phys. Rev. B},
  volume = {97},
  issue = {9},
  pages = {094110},
  numpages = {15},
  year = {2018},
  month = {Mar},
  publisher = {American Physical Society}
}

@article{Sedlmayr2018fate,
  title = {Fate of dynamical phase transitions at finite temperatures and in open systems},
  author = {Sedlmayr, N. and Fleischhauer, M. and Sirker, J.},
  journal = {Phys. Rev. B},
  volume = {97},
  issue = {4},
  pages = {045147},
  numpages = {8},
  year = {2018},
  month = {Jan},
  publisher = {American Physical Society}
}

@article{Bhattacharya2017mixed,
  title = {Mixed state dynamical quantum phase transitions},
  author = {Bhattacharya, Utso and Bandyopadhyay, Souvik and Dutta, Amit},
  journal = {Phys. Rev. B},
  volume = {96},
  issue = {18},
  pages = {180303},
  numpages = {5},
  year = {2017},
  month = {Nov},
  publisher = {American Physical Society}
}

@article{Heyl2017dynamical,
  title = {Dynamical topological quantum phase transitions for mixed states},
  author = {Heyl, M. and Budich, J. C.},
  journal = {Phys. Rev. B},
  volume = {96},
  issue = {18},
  pages = {180304},
  numpages = {5},
  year = {2017},
  month = {Nov},
  publisher = {American Physical Society}
}

@article{Lang2018concurrence,
  title = {Concurrence of dynamical phase transitions at finite temperature in the fully connected transverse-field Ising model},
  author = {Lang, Johannes and Frank, Bernhard and Halimeh, Jad C.},
  journal = {Phys. Rev. B},
  volume = {97},
  issue = {17},
  pages = {174401},
  numpages = {15},
  year = {2018},
  month = {May},
  publisher = {American Physical Society}
}

@article{Kyaw2020dynamical,
  title = {Dynamical quantum phase transitions and non-{Markovian} dynamics},
  author = {Kyaw, Thi Ha and Bastidas, Victor M. and Tangpanitanon, Jirawat and Romero, Guillermo and Kwek, Leong-Chuan},
  journal = {Phys. Rev. A},
  volume = {101},
  issue = {1},
  pages = {012111},
  numpages = {10},
  year = {2020},
  month = {Jan},
  publisher = {American Physical Society}
}

@article{Naji2022dissipative,
  title = {Dissipative Floquet dynamical quantum phase transition},
  author = {Naji, J. and Jafari, Masoud and Jafari, R. and Akbari, Alireza},
  journal = {Phys. Rev. A},
  volume = {105},
  issue = {2},
  pages = {022220},
  numpages = {11},
  year = {2022},
  month = {Feb},
  publisher = {American Physical Society}
}

@article{Kawabata2023dynamical,
  title = {Dynamical quantum phase transitions in {Sachdev-Ye-Kitaev Lindbladians}},
  author = {Kawabata, Kohei and Kulkarni, Anish and Li, Jiachen and Numasawa, Tokiro and Ryu, Shinsei},
  journal = {Phys. Rev. B},
  volume = {108},
  issue = {7},
  pages = {075110},
  numpages = {14},
  year = {2023},
  month = {Aug},
  publisher = {American Physical Society}
}

@article{Xiao2024dynamical,
  title = {Dynamical topological quantum phase transitions in high-order topological systems},
  author = {Xiao, Hai-Xiao and Jiang, Weilun and Qian, Peng and Li, Hongju and Li, Zhongjun and Shen, Heng and Chen, Bing},
  journal = {Phys. Rev. B},
  volume = {110},
  issue = {6},
  pages = {064306},
  numpages = {8},
  year = {2024},
  month = {Aug},
  publisher = {American Physical Society}
}

@article{Maslowski2023dynamical,
  title = {Dynamical bulk-boundary correspondence and dynamical quantum phase transitions in higher-order topological insulators},
  author = {Mas\l{}owski, T. and Sedlmayr, N.},
  journal = {Phys. Rev. B},
  volume = {108},
  issue = {9},
  pages = {094306},
  numpages = {13},
  year = {2023},
  month = {Sep},
  publisher = {American Physical Society}
}

@article{Ye2025disentangling,
  title = {Disentangling connection between static and dynamical phase transitions},
  author = {Ye, Shihao and Khan, Niaz Ali and Sajid, Muhammad},
  journal = {Phys. Rev. A},
  volume = {111},
  issue = {4},
  pages = {042208},
  numpages = {9},
  year = {2025},
  month = {Apr},
  publisher = {American Physical Society}
}

@article{Poyhonen2021entanglement,
  title = {Entanglement echo and dynamical entanglement transitions},
  author = {P\"oyh\"onen, Kim and Ojanen, Teemu},
  journal = {Phys. Rev. Res.},
  volume = {3},
  issue = {4},
  pages = {L042027},
  numpages = {6},
  year = {2021},
  month = {Nov},
  publisher = {American Physical Society}
}

@article{Abeling2016quantum,
  title = {Quantum quench dynamics in the transverse field Ising model at nonzero temperatures},
  author = {Abeling, Nils O. and Kehrein, Stefan},
  journal = {Phys. Rev. B},
  volume = {93},
  issue = {10},
  pages = {104302},
  numpages = {10},
  year = {2016},
  month = {Mar},
  publisher = {American Physical Society}
}

@article{Budich2016dynamical,
  title = {Dynamical topological order parameters far from equilibrium},
  author = {Budich, Jan Carl and Heyl, Markus},
  journal = {Phys. Rev. B},
  volume = {93},
  issue = {8},
  pages = {085416},
  numpages = {7},
  year = {2016},
  month = {Feb},
  publisher = {American Physical Society}
}

@article{Chen2025dynamical,
  title={Dynamical quantum phase transition with divergent multipartite entanglement},
  author={Chen, Jie and de Almeida, Ricardo Costa and Weimer, Hendrik},
  journal={arXiv:2506.13898},
  year={2025}
}

@article{Heyl2014dynamical,
  title = {Dynamical Quantum Phase Transitions in Systems with Broken-Symmetry Phases},
  author = {Heyl, M.},
  journal = {Phys. Rev. Lett.},
  volume = {113},
  issue = {20},
  pages = {205701},
  numpages = {5},
  year = {2014},
  month = {Nov},
  publisher = {American Physical Society}
}

@article{Gilles2025reduced,
  title = {Reduced fidelities for free fermions out of equilibrium: {From} dynamical quantum phase transitions to {Mpemba} effect},
  author = {Gilles Parez, Vincenzo Alba},
  journal={arXiv:2509.01608},
  year={2025}
}

@article{Yang2017dynamical,
  title = {Dynamical signature of localization-delocalization transition in a one-dimensional incommensurate lattice},
  author = {Yang, Chao and Wang, Yucheng and Wang, Pei and Gao, Xianlong and Chen, Shu},
  journal = {Phys. Rev. B},
  volume = {95},
  issue = {18},
  pages = {184201},
  numpages = {6},
  year = {2017},
  month = {May},
  publisher = {American Physical Society}
}

@article{Ye2024energy,
  title = {Energy-dependent dynamical quantum phase transitions in quasicrystals},
  author = {Ye, Shihao and Zhou, Ziheng and Khan, Niaz Ali and Xianlong, Gao},
  journal = {Phys. Rev. A},
  volume = {109},
  issue = {4},
  pages = {043319},
  numpages = {11},
  year = {2024},
  month = {Apr},
  publisher = {American Physical Society}
}

@article{Lang2018dyn,
  title = {Dynamical quantum phase transition for mixed states in open systems},
  author = {Lang, Haifeng and Chen, Yixin and Hong, Qiantan and Fan, Heng},
  journal = {Phys. Rev. B},
  volume = {98},
  issue = {13},
  pages = {134310},
  numpages = {7},
  year = {2018},
  month = {Oct},
  publisher = {American Physical Society}
}

@article{Landau1932theorie,
  title={Zur theorie der energieubertragung. II},
  author={Landau, Lev},
  journal={Physikalische Zeitschrift der Sowjetunion},
  volume={2},
  pages={46},
  year={1932}
}

@article{Stueckelberg1932theorie,
  title={Theorie der unelastischen St{\"o}sse zwischen Atomen},
  author={Stueckelberg, Ernst Carl Gerlach},
  journal={Helv. Phys. Acta},
  volume={5},
  pages={369},
  year={1932}
}

@article{Jordan1928paulische,
  title={{\"U}ber das paulische {\"a}quivalenzverbot},
  author={Jordan, Pascual and Wigner, Eugene},
  journal={Z. Phys.},
  volume={47},
  number={9},
  pages={631--651},
  year={1928},
  publisher={Springer}
}

@article{Andraschko2014dyn,
  title = {Dynamical quantum phase transitions and the {Loschmidt echo: A transfer} matrix approach},
  author = {Andraschko, F. and Sirker, J.},
  journal = {Phys. Rev. B},
  volume = {89},
  issue = {12},
  pages = {125120},
  numpages = {13},
  year = {2014},
  month = {Mar},
  publisher = {American Physical Society}
}

@article{Karrasch2017dyn,
  title = {Dynamical quantum phase transitions in the quantum Potts chain},
  author = {Karrasch, C. and Schuricht, D.},
  journal = {Phys. Rev. B},
  volume = {95},
  issue = {7},
  pages = {075143},
  numpages = {4},
  year = {2017},
  month = {Feb},
  publisher = {American Physical Society}
}

\vspace{0.5cm}

\end{document}